%% file: 0-main.tex
\documentclass[sigconf,numbers]{acmart}
\usepackage{textcomp}
\AtBeginDocument{%
  }

\usepackage{tcolorbox}
\usepackage{xcolor}
\usepackage[normalem]{ulem}

\usepackage{xspace}
\usepackage{hyperref}
\usepackage{multirow}
\usepackage{amsmath}

\usepackage{amssymb}
\usepackage{enumitem} 

\begin{document}

\title{Can QPP Choose the Right Query Variant? \\ Evaluating Query Variant Selection for RAG Pipelines}

\author{Negar Arabzadeh}
 \orcid{0000-0002-4411-7089}
 \affiliation{%
   \institution{UC Berkeley}
  \city{Berkeley}
  \state{CA}
  \country{United States}
   }

\author{Andrew Drozdov}
 \orcid{0000-0002-1025-5715}
 \affiliation{%
   \institution{Databricks}
  \city{San Francisco}
  \state{CA}
  \country{United States}
   }

\author{Michael Bendersky}
 \orcid{0000-0002-2941-6240}
 \affiliation{%
   \institution{Databricks}
  \city{San Francisco}
  \state{CA}
  \country{United States}
   }

\author{Matei Zaharia}
 \orcid{0000-0002-7547-7204}
 \affiliation{%
   \institution{UC Berkeley}
  \city{Berkeley}
  \state{CA}
  \country{United States}
   }
   



\begin{abstract}
  \input{abstract}
\end{abstract}


\copyrightyear{2026}
\acmYear{2026}
\setcopyright{cc}
\setcctype{by}
\acmConference[SIGIR '26]{Proceedings of the 49th International ACM SIGIR Conference on Research and Development in Information Retrieval}{July 20--24, 2026}{Melbourne, VIC, Australia}
\acmBooktitle{Proceedings of the 49th International ACM SIGIR Conference on Research and Development in Information Retrieval (SIGIR '26), July 20--24, 2026, Melbourne, VIC, Australia}
\acmDOI{10.1145/3805712.3808571}
\acmISBN{979-8-4007-2599-9/2026/07}
\maketitle
\input{1-introduction}
\input{2-related-works}
\input{3-experimentalSetup}
\input{4-results}
\input{5-discussion}
\input{6-conclusion}


\bibliographystyle{ACM-Reference-Format}
\bibliography{sample-base}

\end{document}

%% file: abstract.tex
Large Language Models (LLMs) have made query reformulation ubiquitous in modern retrieval and Retrieval-Augmented Generation (RAG) pipelines, enabling the generation of multiple semantically equivalent query variants. However, executing the full pipeline for every reformulation is computationally expensive, motivating selective execution: can we identify the best query variant before incurring downstream retrieval and generation costs?
We investigate Query Performance Prediction (QPP) as a mechanism for variant selection across  ad-hoc retrieval,  and end-to-end RAG. Unlike traditional QPP, which estimates query difficulty across topics, we study intra-topic discrimination—selecting the optimal reformulation among competing variants of the same information need.
Through large-scale experiments on TREC-RAG using both sparse and dense retrievers, we evaluate pre- and post-retrieval predictors under correlation- and decision-based metrics. Our results reveal a systematic divergence between retrieval and generation objectives: variants that maximize ranking metrics such as nDCG often fail to produce the best generated answers, exposing a “utility gap” between retrieval relevance and generation fidelity.
Nevertheless, QPP can reliably identify variants that improve end-to-end quality over the original query. Notably, lightweight pre-retrieval predictors frequently match or outperform more expensive post-retrieval methods, offering a latency-efficient approach to robust RAG.

%% file: 1-introduction.tex
\section{Introduction}

Retrieval-Augmented Generation (RAG) has rapidly become a dominant architectural paradigm for modern information systems \cite{gao2023retrieval,singh2025agentic,zhao2024longrag}. Unlike traditional ad-hoc retrieval, where users directly consume a ranked list, RAG inserts a Large Language Model (LLM) between retrieval and the user, delegating answer synthesis to a generative model conditioned on retrieved evidence \cite{lewis2020retrieval}. This shift fundamentally alters both the objective and the economics of search \cite{krishna2025fact,petroni2024ir}.
In this setting, \textit{query formulation} plays an amplified role \cite{killingback2025benchmarking,chan2024rq,ma2023query}. A user’s original query may fail to retrieve passages that adequately ground generation, exacerbating vocabulary mismatch, intent drift, and underspecification \cite{furnas1987vocabulary,voorhees1993using,miutra2018introduction}. LLM-based query reformulation has become common practice to mitigate this problem by generating multiple semantically equivalent query variants to improve recall and coverage before answer synthesis \cite{dhole2024genqrensemble,li2023pseudo,ye2023enhancing,hosseini2024enhanced,bigdeli2024learning}.

One straightforward strategy is to over-generate query variants, execute the full pipeline for each, and then select the best final answer. A system may generate multiple semantically equivalent reformulations, retrieve and synthesize an answer for each, and use an LLM-as-a-judge, majority voting, or answer-level scoring to choose the most faithful response \cite{wang2022self,ray2025metis,asai2024self}. While this “generate-all-then-select” paradigm directly optimizes answer quality, it is computationally expensive. Although generating query variants is cheap, executing retrieval, reranking, and long-context generation for every variant requires multiple full LLM inference passes. The cost scales linearly with the number of reformulations, increasing latency and monetary expense. In production settings, such exhaustive execution is often infeasible. This motivates a more efficient alternative: \textit{can we identify the most promising query variant before incurring the downstream generation cost?}

Query Performance Prediction (QPP) offers a natural mechanism for this problem \cite{carmel2010estimating,carmel2006makes,arabzadeh2024query,arabzadeh2025vap3}. Traditionally, QPP estimates retrieval effectiveness without relevance judgments and has been used for tasks such as selective query expansion, system routing, and risk-sensitive retrieval \cite{meng2023query,meng2025query,cronen2002predicting}. Its evaluation has largely relied on correlation with ranking metrics such as nDCG or Average Precision.
While correlation remains the standard criterion, it measures statistical association rather than downstream decision quality. As shown by \citet{ganguly2022analysis}, QPP effectiveness varies across evaluation settings—retrieval models, metrics, and rank cutoffs—raising reproducibility concerns and highlighting limitations of correlation-based evaluation.

Recent work broadens QPP evaluation beyond the single-query, single-ranker setting. \citet{DBLP:conf/sigir/ZendelSRKC19} shows that standard formulations conflate queries with the underlying information need and that prediction quality depends on how well a query represents that need, motivating multiple query representations. Building on this, Santra et al. \cite{santra2026breakingflatgeneralisedquery,santra2026beyond} proposes generalized, downstream-aware frameworks that evaluate QPP by its ability to support ranker selection and improve retrieval in fusion pipelines, rather than solely by correlation. Collectively, these studies suggest that statistical agreement alone is insufficient to capture QPP’s practical utility.
More directly related to query reformulation, \citet{scells2018query} introduced Query Variation Performance Prediction (QVPP) in the CLEF 2017 TAR task~\cite{kanoulas2018clef}, aiming to select the most effective manually crafted Boolean query variant in a high-recall medical setting. While this demonstrates the feasibility of variant-level prediction, it is restricted to Boolean retrieval in a specific domain and does not consider neural retrievers, LLM-generated reformulations, or generation-based tasks \cite{bigdeli2026reformer}. In contrast, recent RAG studies show that retrieval effectiveness and downstream utility are weakly aligned~\cite{tian2025relevance,bigdeli2024evaluating,arabzadeh2024adapting}, and that predicted retrieval quality can guide adaptive retrieval in agentic RAG systems~\cite{tian2025right}. Building on this view of QPP as a decision-support mechanism, we investigate whether QPP can select among over-generated query variants prior to retrieval or generation.

Applying QPP to RAG, however, introduces two fundamental shifts. First, the objective changes: classical QPP predicts \textit{retrieval relevance}, whereas RAG ultimately optimizes \textit{generation utility}, i.e., the ability of retrieved evidence to support a faithful and complete answer. Recent work highlights a “utility gap,” where documents that score highly under ranking metrics do not necessarily improve generation quality~\cite{tian2025relevance,salemi2024evaluating,arabzadeh2024comparison,alaofi2024generative}. Second, the decision setting differs. Traditional QPP primarily estimates difficulty across different information needs (\textit{inter-topic prediction}), whereas query reformulation produces multiple variants of the same information need. The task, therefore, becomes one of \textit{intra-topic discrimination} in selecting the single most promising variant among competing reformulations.

Despite these advances, existing work remains largely retrieval-centric, focusing either on estimating query difficulty or selecting among rankers. In contrast, we study QPP in the context of variant selection over LLM-generated reformulations, where the objective is not merely to predict retrieval effectiveness but to identify the variant that yields the strongest end-to-end performance.
To this end, we consider two information access paradigms: \emph{(i) retrieval-only}, where output quality is measured using ranking metrics; and \emph{(ii) retrieval-augmented generation}, where retrieved documents condition the generation process, and performance is measured using nugget-based utility. We evaluate QPP from a utility-oriented perspective across these two paradigms. This framing allows us to examine whether retrieval-optimal variants align with RAG-optimal ones and to quantify the extent of divergence between ranking-based and generation-based objectives.
To structure our investigation, we address the following research questions:

\begin{itemize}[leftmargin=*, itemsep=2pt, topsep=2pt]
\item \textbf{RQ1:} Can pre-retrieval QPP methods reliably identify the best-performing query variant for retrieval and RAG?

\item \textbf{RQ2:} Do post-retrieval QPP methods provide additional gains in variant selection accuracy for retrieval and RAG?

\item \textbf{RQ3:} Does strong retrieval performance prediction translate into strong end-to-end RAG performance prediction?

\item \textbf{RQ4:} Does correlation-based evaluation capture QPP’s utility in query variant selection, or should a more utility-oriented evaluation be preferred?
\end{itemize}

Through a large-scale replicability study on the TREC-RAG 2024 benchmark, we re-implement and evaluate a comprehensive suite of established pre- and post-retrieval QPP baselines within a unified experimental framework~\cite{pradeep2024initial,pradeep2025ragnarok,thakur2025assessing}. For each information need, we generate 30 LLM-based query variants and evaluate them using a broad spectrum of QPP methods, ranging from traditional lexical and statistical predictors to more recent supervised, neural, and transformer-based approaches. We systematically apply these predictors across all query variants, evaluate both sparse and dense retrievers, and  assess performance under two information-seeking paradigms: retrieval-only and RAG.

Our results provide a systematic comparison of QPP behavior across these settings. While QPP methods can identify variants that improve over the original query in certain cases, we observe that strong retrieval prediction does not consistently translate into improved end-to-end generation quality. By releasing all code, query variants, retriever configurations, and evaluation scripts, we aim to establish a transparent and extensible benchmark for studying QPP in modern retrieval and generation pipelines.
We release all data and code at \url{https://github.com/Narabzad/QPP-4-RAG}.

%% file: 2-related-works.tex
\vspace{-1em}
\section{Related Work}

\textbf{QPP Methods.}
QPP estimates query effectiveness without relevance judgments and is commonly categorized into pre-retrieval and post-retrieval methods \cite{carmel2010estimating,arabzadeh2025query,arabzadeh2024query}.

\textit{Pre-retrieval Methods.}
Pre-retrieval predictors rely solely on query and collection statistics. Classical approaches leverage term specificity and distributional signals such as IDF variants, SCS, SCQ, and ICTF~\cite{zhao2008effective,hauff2010predicting,kwok1996new}. More recent methods incorporate semantic representations, including embedding-based predictors and supervised transformer models that regress effectiveness directly from query text~\cite{salamat2023neural,salamat2025contrastive,khodabakhsh2024bertpe,arabzadeh2020neural,saleminezhad2024context}. State-of-the-art approaches such as Query Space Distance (QSD) assume a smooth performance landscape in semantic space, estimating effectiveness by interpolating from neighboring queries in embedding space \cite{amin}.

\textit{Post-retrieval Methods.}
Post-retrieval predictors analyze properties of the retrieved documents and often achieve higher predictive power at the higher computational cost \cite{arabzadeh2023noisy,salamat2023neural,arabzadeh2021query}. Classical score-distribution methods, including Clarity, WIG, NQC, and SMV and more, remain strong baselines~\cite{zhou2007query,shtok2012predicting,cummins2011improved,shtok2010using}. These approaches measure signals such as the KL-divergence between the query-induced language model and the retrieved documents, or the variance of top retrieval scores, under the intuition that well-separated score distributions indicate easier queries.
More recently, supervised QPP models directly learn to predict performance by fine-tuning neural models conditioned on retrieval outputs, often achieving SOTA results for specific retrievers \cite{hashemi2019performance,bertqpp,datta2022pointwise,ebrahimi2024estimating}. Generative relevance estimation methods further extend this paradigm by using LLMs to estimate document utility and approximate ranking metrics \cite{meng2025query,saleminezhad2026structure,saleminezhad2026learning}. While effective, these approaches are computationally expensive.

\textbf{Evaluation and Applications of QPP.}
Traditional QPP evaluation relies on correlation with ranking metrics. However, correlation alone may not reflect practical utility. Recent work advocates decision-oriented evaluation, assessing how well QPP supports downstream tasks such as ranker selection, routing, or risk-sensitive retrieval~\cite{santra2026beyond,santra2026breakingflatgeneralisedquery,roitman2019study}. \citet{ganguly2022analysis} further highlights the instability and context sensitivity of correlation-based assessment, emphasizing the need to evaluate QPP with respect to task-level outcomes—especially when retrieval is part of a larger pipeline.
Consistent with this decision perspective, QPP has been widely used as a control signal in retrieval systems, informing selective query expansion, fusion weighting, and per-query routing between sparse and dense retrievers \cite{arabzadeh2022unsupervised,chifu2025uncovering,faggioli2023geometric}. More recently, QPP-inspired signals have been explored in RAG pipelines to determine when retrieval is necessary, while parallel work in prompt performance prediction highlights performance variance across semantically equivalent inputs \cite{tian2025relevance,tian2025right}.

\textbf{Our Position}
Prior work largely frames QPP as a retrieval-centric difficulty estimator or system selection signal. We instead study QPP as a generation-aware mechanism for selecting among LLM-generated query variants, explicitly evaluating its impact on retrieval and RAG. This reframing enables us to examine the divergence between retrieval-optimal and generation-optimal variants within a unified framework.

%% file: 3-experimentalSetup.tex
\section{Experimental Setup}

\subsection{Dataset}

To study whether QPP can reliably select the best query variant for both retrieval and RAG pipelines, we require a benchmark that supports evaluation at multiple stages of the pipeline. 
Specifically, the dataset must allow separate assessment of (i) retrieval effectiveness and (ii) end-to-end RAG performance.
For this reason, we conduct our experiments on the TREC-RAG 2024 benchmark. 
TREC-RAG is designed explicitly to evaluate RAG systems and provides evaluation protocols for retrieval and RAG tasks separately. 
The benchmark consists of 56 queries constructed over the MS MARCO v2.1 corpus, which contains over 138 Million passages. 
Importantly, these queries have been carefully and thoroughly judged across retrieval and generative dimensions by both human assessors and LLM-based judges, enabling a fair comparison of performance under different pipeline configurations. In this work, we specifically utilize human annotations for both retrieval and nugget-based evaluations, as detailed further in Section \ref{sec:sys_perf}.

\subsection{Query Reformulation}

To generate diverse query variants corresponding to the same underlying information need, we adopt a range of state-of-the-art LLM-based reformulation methods implemented through the {QueryGym} framework \cite{bigdeli2025querygym} including:

\begin{itemize}[leftmargin=*]
    \item \textbf{Generative Query Reformulation (GenQR)}~\cite{wang2023generative}: A zero-shot LLM-based method that directly generates enriched query variants to improve ad-hoc search.

    \item \textbf{GenQR-Ensemble}~\cite{dhole2024genqrensemble}: A GenQR extension that applies an ensemble prompting approach that combines multiple LLM-generated reformulations to create a more robust query variant.
    
    \item \textbf{MuGI} ~\cite{zhang2024exploring}: Generates auxiliary questions and multi-step reasoning expansions to capture latent user intent more effectively.
    
    \item \textbf{QA-Expand }~\cite{seo2025newqueryexpansionapproach}: An expansion strategy that enhances the query by generating multi-question answer pairs to provide broader semantic context.
    
    \item \textbf{Query2Doc} ~\cite{wang2023query2doc}: A generative document-centric strategy where the LLM generates a synthetic {pseudo-document} (answer) that serves to provide a rich set of semantically related terms likely to appear in relevant documents.
    
    \item \textbf{Query2Exp} ~\cite{jagerman2023query}: An expansion-based approach where the LLM generates {explanatory notes} or definitions to clarify the underlying concepts and append this context to the original prompt.
\end{itemize}

All query variants are generated using \texttt{GPT-4o} following the original prompts implemented in the {QueryGym} framework \footnote{\url{https://github.com/ls3-lab/QueryGym}}. We set the temperature to $0.6$ and generated $5$ samples per method to capture semantic diversity while managing variance. This results in $30$ generated variants per information need plus the original query, totaling $31$ variations. 
\subsection{Answer Synthesis}

\paragraph{Retrieval Setup.}
We evaluate query variants using both sparse and dense retrievers to test whether QPP-based selection generalizes across retrieval paradigms.

\begin{itemize}[leftmargin=*]
    \item \textbf{Sparse Retrieval:} BM25 with standard MS MARCO parameters ($k_1=0.9, b=0.4$) via Pyserini. For each query variant, we retrieve the top-100 passages from MS MARCO v2.1.
    
    \item \textbf{Dense Retrieval:} Cohere embeddings with DiskVectorIndex to retrieve the top-100 passages using a compressed, memory-mapped index\footnote{\url{https://huggingface.co/datasets/Cohere/msmarco-v2-embed-english-v3}}.
\end{itemize}

Both configurations were official TREC RAG 2024 baselines. All retrieval parameters are fixed across experiments to isolate the effect of query reformulation.

\paragraph{RAG Setup.}
For RAG, we follow the official TREC-RAG 2024 baseline using the Ragnarok framework \cite{pradeep2025ragnarok}. For each query variant, we retrieve the top-5 documents (sparse or dense) and provide them as context to the generator, which produces an answer conditioned on both the query and the retrieved documents.
We avoid additional prompt engineering to ensure differences arise from query reformulation rather than prompt design. 


\subsection{System Performance Evaluation}
\label{sec:sys_perf}
We first  evaluate the performance of each system configuration under retrieval-only and RAG settings using various query variants. 

\subsubsection{Retrieval Evaluation.}
Since RAG conditions on the top-5 retrieved documents, we report nDCG@5 to align retrieval evaluation with the evidence actually used for generation. We additionally report Recall@100 to capture higher-depth coverage. All metrics are computed using graded relevance judgments on a 0--3 scale provided by human assessors.

\subsubsection{RAG Evaluation.} 
We adopt the official nugget-based evaluation framework from the TREC-RAG 2024 benchmark \cite{pradeep2024initial}. Unlike standard ranking metrics, this framework quantifies how well a synthesized answer satisfies the underlying information need by identifying atomic factual units, or ``nuggets,'' within the response. 
For each query, nuggets are manually created by human assessors based on highly relevant judged documents. Each nugget represents a distinct aspect of the information need and is assigned an importance level: \textsc{Vital} (essential facts) or \textsc{Okay} (supplementary details). System performance is measured by the degree of support provided for each nugget, i.e., Full Support, Partial Support, or No Support.
These support levels are aggregated into weighted utility measures that provide flexibility in prioritizing critical query aspects. We report performance across two ends of the spectrum:
\begin{itemize}[leftmargin=*, nosep]
    \item \textbf{Nugget-All ($\mathbb{N}_{\text{All}}$):} This represents the lenient end of the evaluation spectrum. In this metric, any satisfied nugget—whether \textsc{Vital} or \textsc{Okay}—contributes to the final score, rewarding systems for broad semantic coverage.
    \item \textbf{Nugget-Strict ($\mathbb{N}_{\text{Strict}}$):} This represents the most stringent evaluation criteria. It focuses exclusively on the satisfaction of \textsc{Vital} nuggets, typically requiring a high level of support (e.g., full support) to contribute to the performance score.
\end{itemize}
For details of metric scoring, we refer readers to \cite{pradeep2024initial}.

\subsection{Prediction Performance Evaluation}

We evaluate the capability of QPP methods to estimate the relative quality of query variants derived from the same underlying information need. For each information need, we generate $30$ distinct query variants.
Each variant is executed through the retrieval and RAG pipelines described previously, generating a comprehensive set of ground-truth performance scores. Specifically, we measure retrieval effectiveness using {nDCG@5} and {Recall@100}, and evaluate answer synthesis quality using nugget-based evaluation metrics, including {Nugget-All} ($\mathbb{N}_{\text{all}}$) and {Nugget-Strict} ($\mathbb{N}_{\text{strict}}$).

\subsubsection{Correlation-based Evaluation}

Traditionally, QPP methods are evaluated by computing the correlation between predicted and actual system performance. Given a query $q$, a system $S$ has true performance $M_s(q)$ under metric $M$, while a QPP method predicts a performance score of $\hat{M}_s(q)$.
Now let $Q_I = {q_1, q_2, \dots, q_n}$ denote the $n$ query variants for information need $I$. For each $q_i \in Q_I$, we obtain predicted scores $\hat{M}_s(q_i)$ and true scores $M_s(q_i)$. Prediction quality is measured by the correlation between $\hat{M}_Q$ and $M_Q$, denoted as ${Corr}\langle\hat{M}_Q, M_Q\rangle$. Both linear and rank-based metrics, such as Pearson's $\rho$ and Kendall's $\tau$, are commonly used, and correlations are aggregated across information needs.

While measuring the correlation between predicted and actual performance is standard practice, we argue that this traditional evaluation framework was originally designed for datasets with highly diverse topics (inter-topic difficulty). Therefore, correlation based evaluation may not be the most representative for the query variant selection task. In a production RAG pipeline, a system architect often only requires the predictor to identify the single best-performing variant to execute. If a QPP method successfully selects the optimal variant, the accuracy of the ranking for the remaining sub-optimal variants is of secondary importance. Therefore, we contend that QPP must be evaluated in a more realistic scenario that prioritizes selection accuracy over global correlation. We propose end-to-end evaluation pipelines for assessing the practicality of QPP when choosing among different query variants.
\subsubsection{End-to-End Evaluation}

We evaluate QPP as a practical mechanism for query variant selection rather than a global ranking tool. For each information need $I_i$, let $Q_i = \{q_{i,1}, \dots, q_{i,n}\}$ denote its set of generated variants. We select the query variant predicted to be most effective by a QPP method as:
\setlength{\abovedisplayskip}{6pt}
\setlength{\belowdisplayskip}{6pt}
\begin{equation}
    q_i^{\textsc{QPP}} = 
\arg\max_{q_{i,j} \in Q_i} \hat{M}_s(q_{i,j}),
\end{equation}

where $\hat{M}_s$ denotes the predicted performance score. The selected variant is then executed through the retrieval or RAG pipeline to measure its true system performance $M_s(q_i^{\textsc{QPP}})$.

For comparison, we define the \textsc{Oracle} variant as:
\setlength{\abovedisplayskip}{6pt}
\setlength{\belowdisplayskip}{6pt}
\begin{equation}
q_i^{\textsc{Oracle}} = 
\arg\max_{q_{i,j} \in Q_i} M_s(q_{i,j}),
\end{equation}

\noindent which represents the maximum achievable performance under perfect selection.
We evaluate QPP based on (i) the improvement over the \textsc{Original Query} and (ii) the remaining gap to the \textsc{Oracle}. This evaluation is strictly precision-focused, as it assesses the system's ability to choose the single best variant without being penalized by the  correlation  of sub-optimal variants within the set.

\subsection{QPP baselines}
We consider a comprehensive set of QPP methods:

\subsubsection{Pre-retrieval QPP Methods}

We include the following pre-retrieval QPP methods as baselines in our experiments. For methods that operate at the query-term level, we aggregate term-level signals using different functions (e.g., average, maximum, sum), which are indicated in parentheses for each method. Methods that operate at the whole-query level do not require such aggregation.

\begin{itemize}[leftmargin=*]
    \item \texttt{IDF}: Aggregates the Inverse Document Frequency of query terms using average, maximum, standard deviation, and sum operations to estimate query specificity \cite{kwok1996new}.
    
    \item \texttt{ICTF}: Inverse Collection Term Frequency, which penalizes terms that are frequent in the collection but rare in individual documents \cite{kwok1996new}.
    
    \item \texttt{SCQ}: Collection Query Similarity, which measures the similarity between the query and the collection language model using TF-IDF statistics \cite{zhao2008effective}.
    
    \item \texttt{SCS}: Simplified Clarity Score estimates query ambiguity by approximating the KL-divergence between the query and collection language models \cite{he2004inferring}. We consider two variants: $\texttt{SCS}_\texttt{apx}$, a lightweight approximation based on query length and average \texttt{ICTF}, and $\texttt{SCS}_\texttt{full}$, the full KL-divergence formulation using query and collection term probabilities.
    
    \item \texttt{QL}: Query Likelihood uses the retrieval score of the query generated by the QL model as a proxy for performance \cite{ponte2017language}.
    
    \item \texttt{DM}: Discounted Matryoshka is a neural pre-retrieval method that estimates difficulty based on the distance between the query embedding and a set of reference vectors. We use \texttt{E5} embeddings to obtain the query vector representation \cite{faggioli2023geometric}.
    
    \item \texttt{QSD (Pre)}: Query Space Distance, which estimates performance by interpolating the effectiveness of semantically similar historical queries in the embedding space \cite{bigdeli2025query}.
\end{itemize}

\subsubsection{Post-retrieval QPP Methods.}
These methods leverage the scores and content of the top-$k$ retrieved documents. 

\begin{itemize}[leftmargin=*]
    \item \texttt{Clarity}: Measures the KL-divergence between the language model of the top-$k$ (k=100) retrieved documents and the collection language model \cite{cronen2002predicting}.
    
    \item \texttt{WIG}: Weighted Information Gain computes the information gain of the top retrieved documents compared to the corpus average, acting as a proxy for result quality \cite{zhou2007query}.
    
    \item \texttt{NQC}: Normalized Query Commitment calculates the standard deviation of retrieval scores in the top-$k$ results, assuming that high variance indicates better discrimination between relevant and non-relevant items \cite{shtok2012predicting}.
    
    \item \texttt{SMV}: Score Magnitude and Variance combines the mean and variance of retrieval scores to capture both the strength and discriminative power of the retrieval signal \cite{tao2014query}.
    
    \item \texttt{$\sigma_{\max}$}: The maximum standard deviation of retrieval scores across ranked-list prefixes, capturing peak score dispersion \cite{perez2010standard}.
    
    \item \texttt{$\sigma_{50\%}$}: A QPP method that is based on standard deviation computed over documents scoring at least 50\% of the top score, forming a variable-length ranked list \cite{cummins2011improved}.
    
    \item \texttt{RSD}: Ranking Score Distribution, which analyzes the decay curve of retrieval scores to predict difficulty \cite{roitman2017robust}.
    
    \item \texttt{BERT-QPP}: A supervised method that uses a cross-encoder or bi-encoder to aggregate query-document interaction signals into a performance prediction \cite{arabzadeh2021bert}.
    
    \item \texttt{QSD (Post)}: Extends \texttt{QSD} by incorporating retrieved document information to refine the selection of historical nearest neighbors for performance estimation \cite{bigdeli2025query}.
\end{itemize}

For score-distribution predictors such as \texttt{NQC}, \texttt{WIG}, and \texttt{SMV}, we include both normalized and non-normalized variants. 
Normalization refers to scaling retrieval scores by collection-level statistics (e.g., dividing by the average document score or applying a standardization factor) to reduce sensitivity to absolute score magnitude and improve comparability across queries or retrieval models.

%% file: 4-results.tex
\section{Results and Findings}
\input{table}

Table~\ref{tab:qpp_full_ndcg5} reports end-to-end performance for {retrieval-only} and {RAG} when query variants are selected under different QPP settings.
For each QPP method, we select, for every information need, the query variant with the highest predicted effectiveness score and execute it under the corresponding evaluation paradigm. We report retrieval effectiveness and nugget-based answer quality.
Pre-retrieval QPP methods are applied to both retrieval-only and RAG settings. Post-retrieval QPP methods are likewise applied to both, as they rely on statistics from the retrieved list.
We use the following metrics:

\begin{itemize}
    \item {Nugget-All} ($\mathbb{N}_{\text{all}}$): Lenient end-to-end utility.
    \item {Nugget-Strict} ($\mathbb{N}_{\text{strict}}$): Strict end-to-end utility.
    \item {nDCG@5}: Retrieval quality at depth 5.
    \item {Recall@100}: Retrieval coverage at depth 100.
\end{itemize}
The first row corresponds to the \textit{Original Query} without reformulation, serving as the baseline for retrieval and RAG performance.
The next blocks present results for \textit{pre- and post-retrieval QPP methods}, where for each method, the variant with the highest predicted score is selected.
Finally, the last rows provide \textit{oracle upper bounds per metric}, representing the maximum achievable performance if the best variant were selected using true scores. We report oracle selection based on ranking metrics (Oracle nDCG, Oracle Recall) and answer-level metrics (Oracle Strict, Oracle All). The gap between each QPP method and the oracle quantifies the remaining room for improvement.
Table~\ref{tab:qpp_full_ndcg5} reports the results of QPP-based selection compared to the original query and oracle upper bounds. \uline{Improvements over the original query} are underlined, and the \textbf{best-performing} method within each section is bold.
This table enables us to examine whether QPP-based selection improves over the original query, whether retrieval gains translate into RAG gains, and how far current methods are from the theoretical ceiling.

\vspace{-1em}
\subsection{RQ1: Can pre-retrieval QPP methods reliably identify optimal query variants?}
We analyze whether pre-retrieval QPP signals are sufficient to select variants that improve retrieval and RAG performance. As shown in Table~\ref{tab:qpp_full_ndcg5}, underlined cells indicate improvements over the original query. We observe that pre-retrieval QPP methods almost consistently improve end-to-end RAG performance compared to the original query, with the exception of certain neural predictors such as \texttt{DM}. 
Across most nugget-based metrics ($\mathbb{N}_{\text{Strict}}$ and $\mathbb{N}_{\text{All}}$), the majority of pre-retrieval methods yield improvements, indicating reliable selection of stronger query variants. For BM25, the best-performing pre-retrieval method ($\texttt{IDF}_{max}$) increases $\mathbb{N}_{\text{All}}$ from 0.2730 to 0.3980 (+45.8\%) and $\mathbb{N}_{\text{Strict}}$ from 0.2270 to 0.3770 (+66.1\%) relative to the original query. 
Even in the dense retrieval setting where ranking is already strong, improvements remain notable. However, while these methods substantially improve nugget-based end-to-end performance, gains in retrieval metrics (e.g., nDCG@5 and Recall@100) are much smaller, especially for dense retrievers. Traditional term-based predictors such as $\texttt{IDF}_{avg}$, $\texttt{SCQ}_{avg}$, and $\texttt{ICTF}_{avg}$ consistently improve answer quality but often fail to improve ranking effectiveness. In the dense case, only \texttt{SCS} variants increase nDCG or Recall, yet nugget gains remain consistent across many methods.

This suggests that improvements in generation quality are not strictly driven by improvements in ranking metrics. Rather, pre-retrieval QPP selects variants that retrieve evidence better aligned with downstream synthesis, even when standard ranking measures show limited change.
Moreover, traditional lexical predictors are not only sufficient but frequently outperform neural pre-retrieval methods in selecting variants that enhance answer quality, indicating that signals useful for RAG variant selection differ from those optimized for retrieval difficulty estimation.

\vspace{0.5em}
\begin{tcolorbox}[boxrule=0.5pt, colback=gray!10, arc=4pt, left=3pt, right=3pt, top=3pt, bottom=3pt, boxsep=0pt]
\textbf{Takeaway RQ1:} Pre-retrieval QPP methods can successfully select query variants that improve end-to-end generation performance over the original query. However, they are less effective at improving intermediate retrieval metrics, particularly for dense retrievers, where gains in answer quality do not necessarily correlate with gains in standard ranking metrics.
\end{tcolorbox}

\subsection{RQ2: Do post-retrieval QPP methods improve variant selection accuracy?}

We investigate whether incorporating ranked-list signals in post-retrieval QPP methods leads to stronger variant ranking and improved end-to-end selection compared to pre-retrieval predictors.
Post-retrieval QPP methods demonstrate stronger improvements in retrieval metrics, particularly nDCG@5 and Recall@100 for dense retrievers, as reflected by the greater number of underlined scores in the right-hand columns of the table compared to pre-retrieval methods. For example, under dense retrieval, nDCG@5 increases from 0.5570 to 0.6010 (+7.9\%) with \texttt{RSD}. However, the corresponding improvement in $\mathbb{N}_{\text{All}}$ is only +2.7\%, which is smaller than the gains achieved by the best pre-retrieval methods. This indicates that stronger ranking prediction does not necessarily translate into proportionally stronger end-to-end answer quality.

This pattern is expected, as post-retrieval predictors directly exploit ranked-list signals (e.g., score distributions, clarity) and are therefore optimized to estimate retrieval effectiveness. While they are more effective at identifying variants that improve ranking quality, this advantage does not consistently result in superior downstream answer synthesis ($\mathbb{N}_{\text{Strict}}$ and $\mathbb{N}_{\text{All}}$) compared to the best pre-retrieval methods.

\input{tablecorr2}
From a practical perspective, this distinction is crucial. If the ultimate goal is end-to-end answer quality, pre-retrieval QPP methods offer a compelling alternative. First, they avoid the latency cost of executing an initial retrieval step before prediction. Second, they achieve competitive, and even in some cases stronger, end-to-end improvements. In fact, comparing pre-retrieval vs. post-retrieval performance suggests that selecting a semantically richer query variant may have a larger impact on generation quality than marginally improving the retrieval ranking itself.
Additionally, comparing neural post-retrieval methods (e.g., \texttt{BERT-\allowbreak QPP}) with non-neural ones (e.g., \texttt{NQC}, \texttt{WIG}) reveals that traditional statistical predictors often perform surprisingly well, suggesting they may offer better generalizability across different retrieval models than supervised neural predictors trained on specific distributions.

\vspace{0.5em}
\begin{tcolorbox}[boxrule=0.5pt, colback=gray!10, arc=4pt, left=3pt, right=3pt, top=3pt, bottom=3pt, boxsep=0pt]
\textbf{Takeaway RQ2:} Post-retrieval QPP methods outperform pre-retrieval methods in improving retrieval ranking metrics, particularly for dense retrievers. However, their advantage in end-to-end RAG performance is modest, suggesting that when the goal is answer quality, the additional latency of post-retrieval prediction may not be justified.
\end{tcolorbox}

\subsection{RQ3: Does strong retrieval prediction translate into strong RAG performance prediction?}
We examine whether choosing, for each information need, the query variant that maximizes retrieval effectiveness also yields the strongest end-to-end RAG performance. In other words, \textit{if we optimize the query variant for retrieval metrics, do we also optimize for answer quality?}

\begin{figure*}[t]
    \centering
    \includegraphics[width=0.9\textwidth]{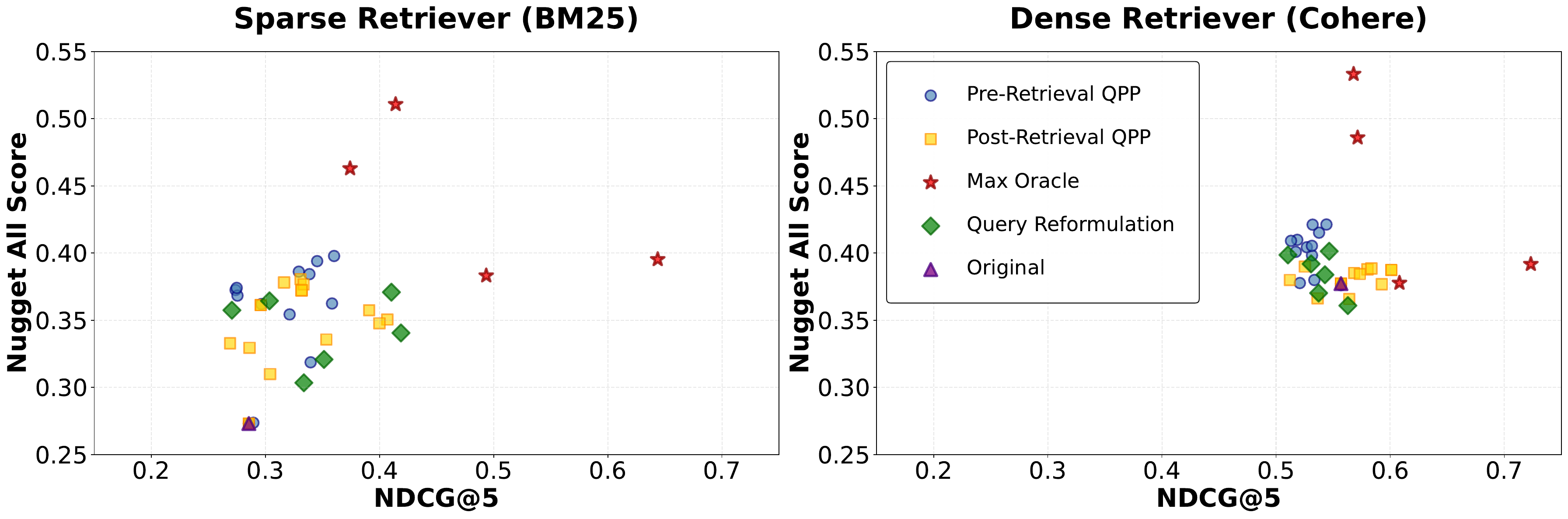}
\caption{Relationship between retrieval effectiveness (nDCG@5) and end-to-end RAG utility (Nugget-All) under sparse and dense retrieval. Each point corresponds to a query variant selected by different strategies (pre-retrieval QPP, post-retrieval QPP, single reformulation, original query, oracle). }

    \label{fig:main}
\end{figure*}

The results in Table~\ref{tab:qpp_full_ndcg5} provide critical insight through the Oracle rows. By comparing oracle selection based on retrieval metrics (e.g., \texttt{Oracle-recall@\allowbreak 100} and \texttt{Oracle-ndcg@\allowbreak 5}) with oracle selection based on nugget metrics (e.g., \texttt{Oracle-}\allowbreak$\mathbb{N}_{\text{All}}$ and \texttt{Oracle-}\allowbreak$\mathbb{N}_{\text{Strict}}$), we observe a substantial divergence between retrieval-optimal and answer-optimal variant selection.

Even in the ideal scenario where, for each information need, we select the query variant that achieves the highest retrieval performance (\texttt{Oracle-ndcg@5}), end-to-end answer quality does not reach its maximum potential. For instance, with the BM25 retriever, nDCG increases from 0.2850 (original query) to 0.6440 under \texttt{Oracle-ndcg@5}. However, the corresponding $\mathbb{N}_{\text{Strict}}$ score for these ``retrieval-optimized'' queries reaches only 0.3440. In contrast, practical QPP methods such as $\texttt{IDF}_{max}$ (pre-retrieval) and \texttt{NQC} (post-retrieval) achieve higher $\mathbb{N}_{\text{Strict}}$ scores of 0.3770 and 0.3550, respectively.
Importantly, this answer quality is not only far from the theoretical maximum, but it is also outperformed by practical QPP methods that do not rely on ground-truth labels. As another example, for dense retrievers, the post-retrieval method \texttt{Clarity} achieves a $\mathbb{N}_{\text{strict}}$ score of 0.3900, which is higher than the score obtained by optimizing directly for \texttt{Oracle-recall@100} (0.3780). This demonstrates that selecting the variant that produces the best-ranked list does not necessarily produce the best final answer.

Conversely, examining oracle selection based on answer quality (\texttt{Oracle-$\mathbb{N}_{\text{Strict}}$}) reveals a dramatically different outcome. Optimizing directly for Strict Nugget yields a score of 0.5360 for BM25 and 0.5690 for the Dense Retriever—substantially higher than the retrieval-optimized selection (0.3440 for BM25 and 0.3540 for Dense when optimized using \texttt{Oracle-ndcg@5}). 
However, these ``answer-optimized'' variants often exhibit lower retrieval effectiveness (e.g., BM25 nDCG decreases from 0.5698 to 0.3480). 
This contrast confirms that the query variant maximizing ranking metrics often differs from the one maximizing end-to-end answer utility. In other words, retrieval-optimal and RAG-optimal selection are fundamentally distinct objectives. Factors that improve ranking quality—such as early precision—do not fully capture downstream answer needs, which may depend on coverage, complementarity, and alignment with the generative model.
Therefore, even perfect retrieval prediction would not guarantee optimal RAG performance if the optimized objective is misaligned with answer quality.

\vspace{-1em}
\begin{tcolorbox}[boxrule=0.5pt, colback=gray!10, arc=4pt, left=3pt, right=3pt, top=3pt, bottom=3pt, boxsep=0pt]
\textbf{Takeaway RQ3:} Selecting the query variant that maximizes retrieval effectiveness does not guarantee optimal end-to-end RAG performance even in ideal scenario. Retrieval-optimized variants yield substantially lower end-to-end quality than answer-optimized or QPP-selected variants, revealing a fundamental misalignment between retrieval-optimal and RAG-optimal selection.
\end{tcolorbox}

\subsection{RQ4: Does correlation-based evaluation capture QPP’s utility in RAG variant selection?}

There has been a longstanding debate in the QPP community regarding the suitability of correlation coefficients as the primary evaluation metric \cite{hauff2009combination,hauff2010query,faggioli2022smare}. Similar to \cite{tian2025right,tian2025relevance}, We extend this discussion to the RAG setting, asking whether strong correlation with retrieval metrics reliably reflects a method’s usefulness for selecting the best-performing query variant in terms of end-to-end answer quality.
Table~\ref{tab:qpp_correlation_formatted} reports the Pearson correlation between QPP predictions and true performance across query variants, measured against both retrieval metrics (nDCG@5, Recall@100) and nugget-based RAG metrics. The results reveal a clear disconnect between these objectives.

For instance, the post-retrieval method \texttt{nqc} applied to the dense retriever exhibits a strong positive correlation with retrieval effectiveness ($r=0.3286$ for nDCG@5, $r=0.4206$ for Recall@100), yet its correlation with end-to-end Nugget scores is negligible ($-0.00384$ for $\mathbb{N}_{\text{All}}$and -0.0145 for $\mathbb{N}_{\text{Strict}}$). This illustrates a critical failure mode that  a QPP method may accurately predict ranking quality while providing no useful signal for selecting variants that improve answer quality.
Conversely, $\texttt{SCQ}_{max}$ in BM25 shows the opposite pattern. It achieves moderate correlation with Nugget scores ($r=0.2436$), despite having near-zero correlation with nDCG@5 ($r=0.0447$). This suggests that the signals governing downstream answer utility differ from those that explain ranking difficulty.

These findings show that accurate retrieval prediction does not guarantee accurate RAG performance prediction. The relationship between ranking quality and answer utility is non-linear, influenced by factors such as complementarity, coverage, redundancy, and the model’s sensitivity to context composition \cite{razavi2025benchmarking}.
More importantly, correlation does not align with the goal of query variant selection. While it captures average association, selection is a decision problem—requiring the identification of the single best variant per query. A method may show reasonable correlation yet still fail to rank the optimal variant correctly.

\vspace{0.5em}
\begin{tcolorbox}[boxrule=0.5pt, colback=gray!10, arc=4pt, left=3pt, right=3pt, top=3pt, bottom=3pt, boxsep=0pt]
\textbf{Takeaway RQ4:} Correlation with retrieval metrics is not a sufficient proxy for evaluating QPP in RAG-based variant selection. A method that predicts ranking quality well may still fail to identify the variant that maximizes end-to-end answer utility. Evaluation should therefore align with the final decision objective i.e., selecting variants that improve answer fidelity, rather than relying solely on intermediate ranking correlations.
\end{tcolorbox}

%% file: table.tex
\begin{table*}[t]
\vspace{-1em}
\centering
\small
\renewcommand{\arraystretch}{1.05}
\begin{tabular}{llcccc|cccc}
\toprule
& & \multicolumn{4}{c}{\textbf{Sparse Retriever (BM25)}} & \multicolumn{4}{c}{\textbf{Dense Retriever (Cohere)}} \\
& & \multicolumn{2}{c}{\textbf{RAG}} & \multicolumn{2}{c}{\textbf{Retrieval}} 
  & \multicolumn{2}{c}{\textbf{RAG}} & \multicolumn{2}{c}{\textbf{Retrieval}} \\
\cmidrule(lr){3-4} \cmidrule(lr){5-6} \cmidrule(lr){7-8} \cmidrule(lr){9-10}
\textbf{Category} & \textbf{Method}
& \textbf{$\mathbb{N}_{\text{All}}$}
& \textbf{$\mathbb{N}_{\text{Strict}}$}
& \textbf{nDCG@5}
& \textbf{Recall@100}
& \textbf{$\mathbb{N}_{\text{All}}$}
& \textbf{$\mathbb{N}_{\text{Strict}}$}
& \textbf{nDCG@5}
& \textbf{Recall@100} \\

\midrule
Original & \texttt{original} & 0.273 & 0.227 & 0.285 & 0.178 & 0.377 & 0.328 & 0.557 & 0.375 \\

\midrule
Pre-retrieval & $\texttt{IDF}_{avg}$ & \underline{0.372} & \underline{0.349} & 0.274 & \underline{0.209} & \underline{0.415} & \underline{0.386} & 0.538 & 0.332 \\
& $\texttt{IDF}_{max}$ & \textbf{\underline{0.398}} & \textbf{\underline{0.377}} & \textbf{\underline{0.360}} & \textbf{\underline{0.239}} & \underline{0.398} & \underline{0.386} & 0.532 & 0.353 \\
& $\texttt{IDF}_{sum}$ & \underline{0.386} & \underline{0.367} & \underline{0.329} & \underline{0.217} & \underline{0.41} & \underline{0.384} & 0.519 & 0.343 \\
& $\texttt{ICTF}_{avg}$ & \underline{0.374} & \underline{0.349} & 0.275 & \underline{0.21} & \textbf{\underline{0.421}} & \textbf{\underline{0.392}} & 0.532 & 0.332 \\

& $\texttt{SCQ}_{avg}$ & \underline{0.368} & \underline{0.341} & 0.275 & \underline{0.21} & \textbf{\underline{0.421}} & \underline{0.391} & 0.544 & 0.337 \\
& $\texttt{SCQ}_{max}$ & \underline{0.362} & \underline{0.338} & \underline{0.358} & \underline{0.213} & \underline{0.378} & \underline{0.366} & 0.521 & 0.339 \\
& $\texttt{SCQ}_{sum}$ & \underline{0.384} & \underline{0.363} & \underline{0.339} & \underline{0.219} & \underline{0.409} & \underline{0.386} & 0.513 & 0.341 \\
& $\texttt{SCS}_\texttt{apx}$ & \underline{0.274} & 0.227 & \underline{0.289} & 0.177 & 0.376 & 0.328 & \textbf{0.557} & \textbf{0.375} \\
& $\texttt{SCS}_\texttt{full}$ & \underline{0.362} & \underline{0.336} & \underline{0.296} & \underline{0.21} & \underline{0.401} & \underline{0.373} & 0.517 & 0.327 \\
& \texttt{QL} & \underline{0.394} & \underline{0.368} & \underline{0.345} & \underline{0.228} & \underline{0.404} & \underline{0.38} & 0.527 & 0.348 \\
& $\texttt{QSD}_\texttt{Pre}$ & \underline{0.354} & \underline{0.333} & \underline{0.321} & \underline{0.199} & \underline{0.405} & \underline{0.384} & 0.531 & 0.34 \\
& \texttt{DM} & 0.273 & 0.227 & 0.285 & 0.178 & 0.377 & 0.328 & \textbf{0.557} & \textbf{0.375} \\

\midrule
Post-retrieval & \texttt{RSD} & \underline{0.372} & \underline{0.345} & \underline{0.332} & \underline{0.22} & \underline{0.387} & 0.328 & \textbf{\underline{0.601}} & \underline{0.386} \\
& \texttt{clarity} & \underline{0.333} & \underline{0.279} & 0.269 & \underline{0.197} & \textbf{\underline{0.39}} & 0.322 & 0.525 & 0.349 \\
& \texttt{NQC} & \textbf{\underline{0.381}} & \textbf{\underline{0.355}} & \underline{0.331} & \underline{0.22} & \underline{0.388} & 0.321 & \underline{0.58} & \underline{0.381} \\
& $\texttt{NQC}_{norm}$ & \underline{0.35} & \underline{0.316} & \textbf{\underline{0.407}} & \textbf{\underline{0.254}} & \underline{0.388} & 0.314 & \underline{0.583} & \underline{0.384} \\
& $\texttt{$\sigma$}_\texttt{max}$ & \underline{0.378} & \underline{0.353} & \underline{0.316} & \underline{0.218} & \underline{0.384} & 0.315 & \underline{0.573} & \underline{0.382} \\
& $\texttt{$\sigma$}_{0.5}$ & \underline{0.377} & \underline{0.352} & \underline{0.333} & \underline{0.215} & 0.377 & 0.328 & 0.557 & 0.375 \\
& \texttt{SMV} & \underline{0.372} & \underline{0.345} & \underline{0.332} & \underline{0.22} & \underline{0.387} & 0.328 & \textbf{\underline{0.601}} & \underline{0.386} \\
& $\texttt{SMV}_{norm}$ & \underline{0.348} & \underline{0.313} & \underline{0.4} & \underline{0.252} & 0.377 & 0.317 & \underline{0.593} & \textbf{\underline{0.387}} \\
& \texttt{WIG} & \underline{0.361} & \underline{0.333} & \underline{0.296} & \underline{0.203} & 0.377 & 0.328 & 0.557 & 0.375 \\
& $\texttt{WIG}_{norm}$ & 0.273 & 0.227 & 0.285 & 0.178 & 0.377 & 0.328 & 0.557 & 0.375 \\
& $\texttt{QSD}_{post}$ & \underline{0.357} & \underline{0.322} & \underline{0.391} & \underline{0.207} & \underline{0.385} & \textbf{\underline{0.343}} & \underline{0.568} & 0.352 \\

& $\texttt{BERTQPP}_{bi-encoder}$ & \underline{0.31} & \underline{0.297} & \underline{0.304} & \underline{0.218} & 0.366 & 0.324 & \underline{0.564} & 0.346 \\
& $\texttt{BERTQPP}_{cross-encoder}$ & \underline{0.336} & \underline{0.32} & \underline{0.353} & \underline{0.215} & 0.366 & 0.328 & 0.536 & 0.361 \\

\midrule
Oracle & \texttt{Oracle-ndcg@5} & \underline{0.395} & \underline{0.344} & \textbf{\underline{0.644}} & \underline{0.257} & \underline{0.392} & \underline{0.354} & \textbf{\underline{0.723}} & 0.374 \\
& \texttt{Oracle-recall@100} & \underline{0.383} & \underline{0.363} & \underline{0.493} & \textbf{\underline{0.333}} & \underline{0.378} & \underline{0.332} & \underline{0.608} & \textbf{\underline{0.444}} \\
& \texttt{Oracle-$\mathbb{N}_{\text{All}}$} & \textbf{\underline{0.511}} & \underline{0.485} & \underline{0.414} & \underline{0.235} & \textbf{\underline{0.533}} & \underline{0.485} & \underline{0.568} & 0.356 \\
& \texttt{Oracle-$\mathbb{N}_{\text{strict}}$} & \underline{0.463} & \textbf{\underline{0.536}} & \underline{0.374} & \underline{0.231} & \underline{0.486} & \textbf{\underline{0.569}} & \underline{0.571} & 0.344 \\
\bottomrule
\end{tabular}
\caption{End-to-end performance of QPP-based query variant selection under retrieval-only and RAG settings. For each method, the highest-scoring variant is executed. Results are reported for BM25 and dense retrievers across ranking (nDCG@5, Recall@100) and nugget-based metrics ($\mathbb{N}_{\text{strict}}$, $\mathbb{N}_{\text{all}}$). Underlined values indicate improvements over the original query; bold denotes the best method per section. Oracle rows represent upper bounds under perfect selection.}
\vspace{-2em}
\label{tab:qpp_full_ndcg5}
\end{table*}

%% file: tablecorr2.tex
\begin{table*}[t]
\vspace{-1em}
\centering
\small
\renewcommand{\arraystretch}{1.05}
\begin{tabular}{llcccc|cccc}
\toprule
& & \multicolumn{4}{c}{\textbf{Sparse Retriever (BM25)}} & \multicolumn{4}{c}{\textbf{Dense Retriever (Cohere)}} \\
& & \multicolumn{2}{c}{\textbf{RAG}} & \multicolumn{2}{c}{\textbf{Retrieval}} 
  & \multicolumn{2}{c}{\textbf{RAG}} & \multicolumn{2}{c}{\textbf{Retrieval}} \\
\cmidrule(lr){3-4} \cmidrule(lr){5-6} \cmidrule(lr){7-8} \cmidrule(lr){9-10}
\textbf{Category} & \textbf{Method}
& \textbf{$\mathbb{N}_{\text{All}}$}
& \textbf{$\mathbb{N}_{\text{Strict}}$}
& \textbf{nDCG@5}
& \textbf{Recall@100}
& \textbf{$\mathbb{N}_{\text{All}}$}
& \textbf{$\mathbb{N}_{\text{Strict}}$}
& \textbf{nDCG@5}
& \textbf{Recall@100} \\

\midrule
Pre-retrieval & $\texttt{IDF}_{avg}$ & 0.205 & 0.131 & -0.137 & -0.138 & \textbf{0.160} & 0.132 & -0.131 & -0.210 \\
& $\texttt{IDF}_{max}$ & 0.243 & \textbf{0.189} & \textbf{0.100} & \textbf{0.118} & 0.059 & 0.058 & 0.008 & -0.058 \\
& $\texttt{IDF}_{sum}$ & 0.244 & 0.167 & 0.006 & 0.076 & 0.102 & 0.094 & -0.066 & -0.119 \\
& $\texttt{ICTF}_{avg} $& 0.227 & 0.153 & -0.113 & -0.099 & 0.154 & \textbf{0.137} & -0.127 & -0.192 \\
& $\texttt{SCQ}_{avg}$ & 0.201 & 0.123 & -0.148 & -0.138 & 0.159 & 0.128 & -0.127 & -0.212 \\
& $\texttt{SCQ}_{max}$ & 0.244 & 0.155 & 0.045 & 0.076 & 0.059 & 0.048 & -0.047 & -0.109 \\
& $\texttt{SCQ}_{sum}$ & 0.244 & 0.168 & 0.010 & 0.081 & 0.101 & 0.094 & -0.063 & -0.115 \\
& $\texttt{SCS}_{apx}$ & -0.232 & -0.181 & -0.086 & -0.183 & -0.036 & -0.053 & 0.011 & \textbf{0.058} \\
& $\texttt{SCS}_{full}$ & 0.168 & 0.089 & -0.093 & -0.065 & 0.115 & 0.079 & -0.151 & -0.212 \\
& \texttt{QL} & \textbf{0.244} & 0.172 & 0.029 & 0.108 & 0.092 & 0.092 & -0.046 & -0.091 \\
& $\texttt{QSD}_{Pre}$ & 0.070 & 0.071 & 0.011 & 0.027 & -0.016 & -0.024 & -0.106 & -0.066 \\
& \texttt{DM}  & -0.041 & -0.011 & 0.099 & 0.060 & -0.073 & -0.058 & 0.027 & 0.056 \\
\midrule
Post-retrieval & \texttt{RSD} & 0.230 & 0.158 & -0.010 & 0.073 & -0.024 & -0.002 & 0.320 & 0.417 \\
& \texttt{clarity} & 0.061 & -0.002 & -0.102 & -0.093 & \textbf{0.046} & \textbf{0.009} & -0.022 & -0.002 \\
& \texttt{NQC} & 0.233 & 0.160 & -0.007 & 0.068 & -0.038 & -0.015 & \textbf{0.329} & \textbf{0.421} \\
& $\texttt{NQC}_{norm}$ & 0.086 & 0.094 & \textbf{0.218} & \textbf{0.325} & -0.023 & -0.003 & 0.291 & 0.370 \\
& $\texttt{$\sigma$}_{max}$ & \textbf{0.234} & 0.161 & -0.005 & 0.068 & -0.063 & -0.036 & 0.290 & 0.372 \\
& $\texttt{$\sigma$}_{0.5}$ & 0.234 & \textbf{0.168} & -0.024 & 0.056 & -0.069 & -0.067 & 0.077 & 0.169 \\
& \texttt{SMV} & 0.230 & 0.158 & -0.010 & 0.073 & -0.024 & -0.002 & 0.320 & 0.417 \\
& $\texttt{SMV}_{norm}$ & 0.085 & 0.093 & 0.199 & 0.324 & -0.012 & 0.005 & 0.290 & 0.380 \\
& \texttt{WIG} & 0.220 & 0.133 & -0.092 & -0.050 & -0.068 & -0.063 & 0.033 & 0.119 \\
& $\texttt{WIG}_{norm}$ & -0.253 & -0.191 & -0.052 & -0.113 & -0.057 & -0.054 & 0.019 & 0.106 \\
& $\texttt{QSD}_{Post}$ & 0.070 & 0.034 & -0.082 & -0.028 & 0.012 & -0.024 & 0.061 & 0.079 \\ 
& $\texttt{BERTQPP}_{bi-encoder}$ & 0.082 & 0.106 & 0.046 & 0.071 & -0.070 & -0.062 & 0.077 & 0.074 \\
& $\texttt{BERTQPP}_{cross-encoder}$ & 0.191 & 0.139 & 0.147 & 0.139 & -0.004 & -0.015 & -0.006 & -0.020 \\

\bottomrule
\end{tabular}
\caption{Pearson correlation of QPP methods with retrieval and RAG metrics. Highest correlation in each section is bolded.}
\vspace{-1em}
\label{tab:qpp_correlation_formatted}
\vspace{-2em}
\end{table*}

%% file: 5-discussion.tex
\section{Discussion}
\input{table_qreform}

A central practical question emerging from our results is how a RAG system should use query reformulation in deployment. If multiple reformulations can be generated for the same information need, should the system simply commit to the single reformulation strategy that performs best on average, or should it perform query-dependent variant selection? 
Table~\ref{tab:summary_comparison} reports the strongest configuration from each category for both sparse (BM25) and dense retrieval. Specifically, for each retriever, we present: (i) the original query, (ii) the best-performing single reformulator, (iii) the strongest pre-retrieval QPP method, (iv) the strongest post-retrieval QPP method, and (v) the Oracle upper bounds. Rather than averaging across methods, Table~\ref{tab:summary_comparison} emphasizes the best achievable performance within each paradigm, enabling a direct comparison between committing to a single reformulation strategy and performing query-dependent variant selection. Figure~\ref{fig:main} complements this analysis by plotting retrieval effectiveness (nDCG@5) against end-to-end RAG utility ($\mathbb{N}_{\text{All}}$) under both sparse and dense retrieval.

\paragraph{Query Variants Selection vs. Single Reformulation.}
Here, we study whether variant selection truly improves over simply committing to the strongest reformulator. In Table~\ref{tab:summary_comparison}, MuGI provides the best overall reformulation baseline for BM25 (averaged across five trials), while in dense retrieval the strongest reformulator similarly establishes a competitive baseline.
QPP-based selection consistently improves over these single-reformulator strategies. For BM25, both the best pre-retrieval and best post-retrieval predictors outperform MuGI, demonstrating that selective routing yields additional gains beyond committing to a single reformulation. In dense retrieval, the improvement over the strongest reformulator is smaller (e.g., $\mathbb{N}_{\text{All}}$ improving from approximately 0.4013 to 0.4152), yet still consistent. This suggests that when a reformulator is already strong, the marginal gain from QPP may be moderate—but still meaningful.

Figure~\ref{fig:main} further clarifies this behavior. Single reformulators appear scattered across the plane, indicating that no fixed strategy dominates across information needs. QPP-based selection shifts configurations upward in $\mathbb{N}_{\text{All}}$, even when nDCG gains are limited. Notably, this improvement does not require consistent increases in retrieval metrics, reinforcing that \emph{RAG Performance Prediction} is distinct from traditional retrieval prediction.

Because both plots share the same scale, we can also directly compare sparse and dense retrievers. Dense retrieval achieves higher overall nDCG, but the points are tightly clustered, making configurations harder to distinguish using ranking metrics alone. In contrast, sparse retrieval exhibits greater dispersion, providing a more discriminative landscape for variant selection.

\paragraph{Oracle Gap and Future Promise.}
Despite the improvements that are achieved by QPP-based selection, a substantial gap remains between current methods and the Oracle upper bounds. The oracle results, computed by selecting the variant with the highest true score under each target metric, reveal that retrieval-optimal and answer-optimal variants frequently differ. 
Importantly, the oracle analysis in Tables~\ref{tab:qpp_correlation_formatted} and \ref{tab:summary_comparison} confirms that the best-performing variant is already present among the generated reformulations. The ceiling is therefore not hypothetical—it is attainable within the existing variant pool. In particular, the gains under $\mathbb{N}_{\text{All}}$ and $\mathbb{N}_{\text{Strict}}$ demonstrate that substantially higher end-to-end performance is achievable if the correct variant can be reliably identified. 

While current QPP methods provide consistent but incremental improvements over single reformulation strategies, the remaining oracle gap highlights significant future potential. This positions QPP as a high-impact prediction problem: even modest advances in generation-aware or utility-aligned predictors could translate into large end-to-end gains without increasing the number of executed variants.

\paragraph{Efficiency and Optimization Objectives.}
Finally, these findings have direct system implications. Executing retrieval and generation for all variants and then selecting the best answer is computationally expensive. QPP shifts this decision upstream, selecting the best \emph{input query} before incurring generation costs. Pre-retrieval predictors are particularly attractive due to their low latency.

More broadly, as illustrated in Figure~\ref{fig:main}, optimizing for retrieval (nDCG) is not equivalent to optimizing for RAG utility. Future work should, therefore, develop generation-aware predictors that treat retrieval and generation as a coupled system, directly estimating answer-grounding potential rather than ranking quality alone.

%% file: table_qreform.tex
\begin{table*}[t]
\centering
\small
\renewcommand{\arraystretch}{1.1}
\vspace{-1em}
\begin{tabular}{llcccc|cccc}
\toprule
& & \multicolumn{4}{c}{\textbf{Sparse Retriever (BM25)}} & \multicolumn{4}{c}{\textbf{Dense Retriever (Cohere)}} \\
& & \multicolumn{2}{c}{\textbf{RAG}} & \multicolumn{2}{c}{\textbf{Retrieval}} 
  & \multicolumn{2}{c}{\textbf{RAG}} & \multicolumn{2}{c}{\textbf{Retrieval}} \\
\cmidrule(lr){3-4} \cmidrule(lr){5-6} \cmidrule(lr){7-8} \cmidrule(lr){9-10}
\textbf{Category} & \textbf{Method}
& \textbf{$\mathbb{N}_{\text{All}}$}
& \textbf{$\mathbb{N}_{\text{Strict}}$}
& \textbf{nDCG@5}
& \textbf{Recall@100}
& \textbf{$\mathbb{N}_{\text{All}}$}
& \textbf{$\mathbb{N}_{\text{Strict}}$}
& \textbf{nDCG@5}
& \textbf{Recall@100} \\

\midrule
Original & \texttt{original} & 0.273 & 0.227 & 0.285 & 0.178 & 0.377 & 0.328 & 0.557 & 0.375 \\
\midrule
\multirow{2}{*}{\shortstack[l]{Best\\Pre-retrieval}}
  &  $\texttt{IDF}_{max}$ & \textbf{{0.398}} & \textbf{{0.377}} & \textbf{{0.360}} & \textbf{{0.239}} & {0.398} & {0.386} & 0.532 & 0.353 \\ 
      & $ \texttt{IDF}_{avg}$ &  0.373 & 0.349 & 0.264 & 0.209 & \textbf{0.415} & 0.386 & 0.510 & 0.332 \\
\midrule
\multirow{2}{*}{\shortstack[l]{Best\\Post-retrieval}}
  &  \texttt{NQC} & \textbf{{0.381}} & \textbf{{0.355}} & {0.331} & {0.22} & {0.388} & 0.321 & {0.58} & {0.381}  \\
       & \texttt{Clarity} & {0.333} & {0.279} & 0.269 & {0.197} & \textbf{{0.39}} & 0.322 & 0.525 & 0.349 \\
\midrule
\multirow{2}{*}{\shortstack[l]{Best Vanilla\\Query Reformulation}}
 & MuGI & \textbf{0.371} & \textbf{0.349} & 0.387 & \textbf{0.254} & 0.384 & 0.364 & 0.535 & 0.364 \\
& GenQR & 0.357 & 0.322 & 0.255 & 0.181 & \textbf{0.401} & \textbf{0.375} & 0.520 & 0.335 \\
\midrule

Oracle & \texttt{Oracle-ndcg@5} & {0.395} & {0.344} & \textbf{{0.644}} & {0.257} & {0.392} & {0.354} & \textbf{{0.723}} & 0.374 \\
& \texttt{Oracle-recall@100} & {0.383} & {0.363} & {0.493} & \textbf{{0.333}} & {0.378} & {0.332} & {0.608} & \textbf{{0.444}} \\
& \texttt{Oracle-$\mathbb{N}_{\text{All}}$} & \textbf{{0.511}} & {0.485} & {0.414} & {0.235} & \textbf{{0.533}} & {0.485} & {0.568} & 0.356 \\
& \texttt{Oracle-$\mathbb{N}_{\text{strict}}$} & {0.463} & \textbf{{0.536}} & {0.374} & {0.231} & {0.486} & \textbf{{0.569}} & {0.571} & 0.344 \\
\bottomrule
\end{tabular}
\caption{Best-performing configuration from each paradigm including Original, pre-retrieval QPP, post-retrieval QPP, single reformulation, and Oracle upper bounds, reported separately for BM25 and dense retrieval. Bold values are carried over from Table~\ref{tab:qpp_full_ndcg5} and denote the best-performing method within their corresponding block in Table \ref{tab:qpp_full_ndcg5}.}
\vspace{-2.5em}
\label{tab:summary_comparison}
\end{table*}

%% file: 6-conclusion.tex
\vspace{-0.5em}
\section{Conclusion}

This work provides a comprehensive and fully reproducible study of QPP for query variant selection in RAG systems. We integrate pre- and post-retrieval QPP methods, multiple reformulation strategies, sparse and dense retrievers, and end-to-end generation evaluation within a unified framework. By standardizing datasets, metrics, retriever configurations, reformulation protocols, and random seeds, we enable controlled comparison across paradigms and isolate the effect of variant selection.

Our findings highlight two key insights. First, variant selection consistently improves robustness over committing to a single reformulation strategy, confirming the value of selective execution in RAG pipelines, though gains remain bounded relative to the Oracle ceiling. Second, retrieval metrics and generation utility are not structurally aligned: variants that optimize ranking effectiveness do not necessarily maximize answer quality, underscoring the need for end-to-end evaluation.

More broadly, our results position QPP as a decision-making component in modern RAG systems rather than merely a predictor of retrieval difficulty. By enabling upstream selection of query variants, QPP can reduce computational cost while maintaining or improving answer quality. However, the gap to oracle performance highlights substantial room for improvement, particularly in designing generation-aware predictors that account for context composition and downstream utility.

%% file: 0-main.bbl

\begin{thebibliography}{87}


\ifx \showCODEN    \undefined \def \showCODEN     #1{\unskip}     \fi
\ifx \showISBNx    \undefined \def \showISBNx     #1{\unskip}     \fi
\ifx \showISBNxiii \undefined \def \showISBNxiii  #1{\unskip}     \fi
\ifx \showISSN     \undefined \def \showISSN      #1{\unskip}     \fi
\ifx \showLCCN     \undefined \def \showLCCN      #1{\unskip}     \fi
\ifx \shownote     \undefined \def \shownote      #1{#1}          \fi
\ifx \showarticletitle \undefined \def \showarticletitle #1{#1}   \fi
\ifx \showURL      \undefined \def \showURL       {\relax}        \fi
\providecommand\bibfield[2]{#2}
\providecommand\bibinfo[2]{#2}
\providecommand\natexlab[1]{#1}
\providecommand\showeprint[2][]{arXiv:#2}

\bibitem[Alaofi et~al\mbox{.}(2024)]%
        {alaofi2024generative}
\bibfield{author}{\bibinfo{person}{Marwah Alaofi}, \bibinfo{person}{Negar Arabzadeh}, \bibinfo{person}{Charles~LA Clarke}, {and} \bibinfo{person}{Mark Sanderson}.} \bibinfo{year}{2024}\natexlab{}.
\newblock \showarticletitle{Generative information retrieval evaluation}.
\newblock In \bibinfo{booktitle}{\emph{Information access in the era of generative ai}}. \bibinfo{publisher}{Springer}, \bibinfo{pages}{135--159}.
\newblock


\bibitem[Arabzadeh and Bagheri(2025)]%
        {arabzadeh2025vap3}
\bibfield{author}{\bibinfo{person}{Negar Arabzadeh} {and} \bibinfo{person}{Ebrahim Bagheri}.} \bibinfo{year}{2025}\natexlab{}.
\newblock \showarticletitle{VAP3: Variation-Aware Prompt Performance Prediction}. In \bibinfo{booktitle}{\emph{Proceedings of the 48th International ACM SIGIR Conference on Research and Development in Information Retrieval}}. \bibinfo{pages}{2794--2799}.
\newblock


\bibitem[Arabzadeh et~al\mbox{.}(2024a)]%
        {arabzadeh2024adapting}
\bibfield{author}{\bibinfo{person}{Negar Arabzadeh}, \bibinfo{person}{Amin Bigdeli}, {and} \bibinfo{person}{Charles~LA Clarke}.} \bibinfo{year}{2024}\natexlab{a}.
\newblock \showarticletitle{Adapting standard retrieval benchmarks to evaluate generated answers}. In \bibinfo{booktitle}{\emph{European Conference on Information Retrieval}}. Springer, \bibinfo{pages}{399--414}.
\newblock


\bibitem[Arabzadeh et~al\mbox{.}(2021a)]%
        {arabzadeh2021query}
\bibfield{author}{\bibinfo{person}{Negar Arabzadeh}, \bibinfo{person}{Amin Bigdeli}, \bibinfo{person}{Morteza Zihayat}, {and} \bibinfo{person}{Ebrahim Bagheri}.} \bibinfo{year}{2021}\natexlab{a}.
\newblock \showarticletitle{Query Performance Prediction Through Retrieval Coherency}. In \bibinfo{booktitle}{\emph{Advances in Information Retrieval: 43rd European Conference on IR Research, ECIR 2021, Virtual Event, March 28--April 1, 2021, Proceedings, Part II 43}}. Springer, \bibinfo{pages}{193--200}.
\newblock


\bibitem[Arabzadeh and Clarke(2024)]%
        {arabzadeh2024comparison}
\bibfield{author}{\bibinfo{person}{Negar Arabzadeh} {and} \bibinfo{person}{Charles~LA Clarke}.} \bibinfo{year}{2024}\natexlab{}.
\newblock \showarticletitle{A comparison of methods for evaluating generative ir}.
\newblock \bibinfo{journal}{\emph{arXiv preprint arXiv:2404.04044}} (\bibinfo{year}{2024}).
\newblock


\bibitem[Arabzadeh et~al\mbox{.}(2023)]%
        {arabzadeh2023noisy}
\bibfield{author}{\bibinfo{person}{Negar Arabzadeh}, \bibinfo{person}{Radin Hamidi~Rad}, \bibinfo{person}{Maryam Khodabakhsh}, {and} \bibinfo{person}{Ebrahim Bagheri}.} \bibinfo{year}{2023}\natexlab{}.
\newblock \showarticletitle{Noisy perturbations for estimating query difficulty in dense retrievers}. In \bibinfo{booktitle}{\emph{Proceedings of the 32nd ACM international conference on information and knowledge management}}. \bibinfo{pages}{3722--3727}.
\newblock


\bibitem[Arabzadeh et~al\mbox{.}(2021b)]%
        {bertqpp}
\bibfield{author}{\bibinfo{person}{Negar Arabzadeh}, \bibinfo{person}{Maryam Khodabakhsh}, {and} \bibinfo{person}{Ebrahim Bagheri}.} \bibinfo{year}{2021}\natexlab{b}.
\newblock \showarticletitle{BERT-QPP: Contextualized Pre-trained transformers for Query Performance Prediction}. In \bibinfo{booktitle}{\emph{Proceedings of the 30th ACM International Conference on Information \& Knowledge Management}} (Virtual Event, Queensland, Australia) \emph{(\bibinfo{series}{CIKM '21})}. \bibinfo{publisher}{Association for Computing Machinery}, \bibinfo{address}{New York, NY, USA}, \bibinfo{pages}{2857–2861}.
\newblock
\showISBNx{9781450384469}
\href{https://doi.org/10.1145/3459637.3482063}{doi:\nolinkurl{10.1145/3459637.3482063}}


\bibitem[Arabzadeh et~al\mbox{.}(2021c)]%
        {arabzadeh2021bert}
\bibfield{author}{\bibinfo{person}{Negar Arabzadeh}, \bibinfo{person}{Maryam Khodabakhsh}, {and} \bibinfo{person}{Ebrahim Bagheri}.} \bibinfo{year}{2021}\natexlab{c}.
\newblock \showarticletitle{BERT-QPP: Contextualized Pre-trained transformers for Query Performance Prediction}. In \bibinfo{booktitle}{\emph{CIKM}}.
\newblock


\bibitem[Arabzadeh et~al\mbox{.}(2024b)]%
        {arabzadeh2024query}
\bibfield{author}{\bibinfo{person}{Negar Arabzadeh}, \bibinfo{person}{Chuan Meng}, \bibinfo{person}{Mohammad Aliannejadi}, {and} \bibinfo{person}{Ebrahim Bagheri}.} \bibinfo{year}{2024}\natexlab{b}.
\newblock \showarticletitle{Query performance prediction: Techniques and applications in modern information retrieval}. In \bibinfo{booktitle}{\emph{Proceedings of the 2024 Annual International ACM SIGIR Conference on Research and Development in Information Retrieval in the Asia Pacific Region}}. \bibinfo{pages}{291--294}.
\newblock


\bibitem[Arabzadeh et~al\mbox{.}(2025)]%
        {arabzadeh2025query}
\bibfield{author}{\bibinfo{person}{Negar Arabzadeh}, \bibinfo{person}{Chuan Meng}, \bibinfo{person}{Mohammad Aliannejadi}, {and} \bibinfo{person}{Ebrahim Bagheri}.} \bibinfo{year}{2025}\natexlab{}.
\newblock \showarticletitle{Query performance prediction: Theory, techniques and applications}. In \bibinfo{booktitle}{\emph{Proceedings of the Eighteenth ACM International Conference on Web Search and Data Mining}}. \bibinfo{pages}{991--994}.
\newblock


\bibitem[Arabzadeh et~al\mbox{.}(2022)]%
        {arabzadeh2022unsupervised}
\bibfield{author}{\bibinfo{person}{Negar Arabzadeh}, \bibinfo{person}{Mahsa Seifikar}, {and} \bibinfo{person}{Charles~LA Clarke}.} \bibinfo{year}{2022}\natexlab{}.
\newblock \showarticletitle{Unsupervised question clarity prediction through retrieved item coherency}. In \bibinfo{booktitle}{\emph{Proceedings of the 31st ACM International Conference on Information \& Knowledge Management}}.
\newblock


\bibitem[Arabzadeh et~al\mbox{.}(2020)]%
        {arabzadeh2020neural}
\bibfield{author}{\bibinfo{person}{Negar Arabzadeh}, \bibinfo{person}{Fattane Zarrinkalam}, \bibinfo{person}{Jelena Jovanovic}, \bibinfo{person}{Feras Al-Obeidat}, {and} \bibinfo{person}{Ebrahim Bagheri}.} \bibinfo{year}{2020}\natexlab{}.
\newblock \showarticletitle{Neural embedding-based specificity metrics for pre-retrieval query performance prediction}.
\newblock \bibinfo{journal}{\emph{Information Processing \& Management}} \bibinfo{volume}{57}, \bibinfo{number}{4} (\bibinfo{year}{2020}), \bibinfo{pages}{102248}.
\newblock
\href{https://doi.org/10.1016/j.ipm.2020.102248}{doi:\nolinkurl{10.1016/j.ipm.2020.102248}}


\bibitem[Asai et~al\mbox{.}(2024)]%
        {asai2024self}
\bibfield{author}{\bibinfo{person}{Akari Asai}, \bibinfo{person}{Zeqiu Wu}, \bibinfo{person}{Yizhong Wang}, \bibinfo{person}{Avirup Sil}, {and} \bibinfo{person}{Hannaneh Hajishirzi}.} \bibinfo{year}{2024}\natexlab{}.
\newblock \showarticletitle{Self-rag: Learning to retrieve, generate, and critique through self-reflection}.
\newblock  (\bibinfo{year}{2024}).
\newblock


\bibitem[Bigdeli et~al\mbox{.}(2024a)]%
        {bigdeli2024learning}
\bibfield{author}{\bibinfo{person}{Amin Bigdeli}, \bibinfo{person}{Negar Arabzadeh}, {and} \bibinfo{person}{Ebrahim Bagheri}.} \bibinfo{year}{2024}\natexlab{a}.
\newblock \showarticletitle{Learning to jointly transform and rank difficult queries}. In \bibinfo{booktitle}{\emph{European Conference on Information Retrieval}}. Springer, \bibinfo{pages}{40--48}.
\newblock


\bibitem[Bigdeli et~al\mbox{.}(2024b)]%
        {bigdeli2024evaluating}
\bibfield{author}{\bibinfo{person}{Amin Bigdeli}, \bibinfo{person}{Negar Arabzadeh}, \bibinfo{person}{Ebrahim Bagheri}, {and} \bibinfo{person}{Charles~LA Clarke}.} \bibinfo{year}{2024}\natexlab{b}.
\newblock \showarticletitle{Evaluating relative retrieval effectiveness with normalized residual gain}. In \bibinfo{booktitle}{\emph{Proceedings of the 2024 Annual International ACM SIGIR Conference on Research and Development in Information Retrieval in the Asia Pacific Region}}. \bibinfo{pages}{64--71}.
\newblock


\bibitem[Bigdeli et~al\mbox{.}(2025a)]%
        {amin}
\bibfield{author}{\bibinfo{person}{Amin Bigdeli}, \bibinfo{person}{Sajad Ebrahimi}, \bibinfo{person}{Negar Arabzadeh}, \bibinfo{person}{Sara Salamat}, \bibinfo{person}{Shirin SeyedSalehi}, \bibinfo{person}{Maryam Khodabakhsh}, \bibinfo{person}{Fattane Zarrinkalam}, {and} \bibinfo{person}{Ebrahim Bagheri}.} \bibinfo{year}{2025}\natexlab{a}.
\newblock \showarticletitle{Query Performance Prediction Using Neural Query Space Proximity}.
\newblock \bibinfo{journal}{\emph{ACM Trans. Intell. Syst. Technol.}} (\bibinfo{date}{Sept.} \bibinfo{year}{2025}).
\newblock
\showISSN{2157-6904}
\href{https://doi.org/10.1145/3762197}{doi:\nolinkurl{10.1145/3762197}}


\bibitem[Bigdeli et~al\mbox{.}(2025b)]%
        {bigdeli2025query}
\bibfield{author}{\bibinfo{person}{Amin Bigdeli}, \bibinfo{person}{Sajad Ebrahimi}, \bibinfo{person}{Negar Arabzadeh}, \bibinfo{person}{Sara Salamat}, \bibinfo{person}{Shirin Seyedsalehi}, \bibinfo{person}{Maryam Khodabakhsh}, \bibinfo{person}{Fattane Zarrinkalam}, {and} \bibinfo{person}{Ebrahim Bagheri}.} \bibinfo{year}{2025}\natexlab{b}.
\newblock \showarticletitle{Query Performance Prediction Using Neural Query Space Proximity}.
\newblock \bibinfo{journal}{\emph{ACM Transactions on Intelligent Systems and Technology}} \bibinfo{volume}{17}, \bibinfo{number}{1} (\bibinfo{year}{2025}), \bibinfo{pages}{1--25}.
\newblock


\bibitem[Bigdeli et~al\mbox{.}(2026)]%
        {bigdeli2026reformer}
\bibfield{author}{\bibinfo{person}{Amin Bigdeli}, \bibinfo{person}{Mert Incesu}, \bibinfo{person}{Negar Arabzadeh}, \bibinfo{person}{Charles~LA Clarke}, {and} \bibinfo{person}{Ebrahim Bagheri}.} \bibinfo{year}{2026}\natexlab{}.
\newblock \showarticletitle{ReFormeR: Learning and Applying Explicit Query Reformulation Patterns}. In \bibinfo{booktitle}{\emph{European Conference on Information Retrieval}}. Springer, \bibinfo{pages}{400--408}.
\newblock


\bibitem[Bigdeli et~al\mbox{.}(2025c)]%
        {bigdeli2025querygym}
\bibfield{author}{\bibinfo{person}{Amin Bigdeli}, \bibinfo{person}{Radin~Hamidi Rad}, \bibinfo{person}{Mert Incesu}, \bibinfo{person}{Negar Arabzadeh}, \bibinfo{person}{Charles~LA Clarke}, {and} \bibinfo{person}{Ebrahim Bagheri}.} \bibinfo{year}{2025}\natexlab{c}.
\newblock \showarticletitle{QueryGym: A Toolkit for Reproducible LLM-Based Query Reformulation}.
\newblock \bibinfo{journal}{\emph{arXiv preprint arXiv:2511.15996}} (\bibinfo{year}{2025}).
\newblock


\bibitem[Carmel and Yom-Tov(2010)]%
        {carmel2010estimating}
\bibfield{author}{\bibinfo{person}{David Carmel} {and} \bibinfo{person}{Elad Yom-Tov}.} \bibinfo{year}{2010}\natexlab{}.
\newblock \bibinfo{booktitle}{\emph{Estimating the Query Difficulty for Information Retrieval}}. \bibinfo{series}{Synthesis Lectures on Information Concepts, Retrieval, and Services}, Vol.~\bibinfo{volume}{2}.
\newblock \bibinfo{publisher}{Morgan \& Claypool Publishers}. 1--89 pages.
\newblock


\bibitem[Carmel et~al\mbox{.}(2006)]%
        {carmel2006makes}
\bibfield{author}{\bibinfo{person}{David Carmel}, \bibinfo{person}{Elad Yom-Tov}, \bibinfo{person}{Adam Darlow}, {and} \bibinfo{person}{Dan Pelleg}.} \bibinfo{year}{2006}\natexlab{}.
\newblock \showarticletitle{What makes a query difficult?}. In \bibinfo{booktitle}{\emph{Proceedings of the 29th annual international ACM SIGIR conference on Research and development in information retrieval}}. \bibinfo{pages}{390--397}.
\newblock


\bibitem[Chan et~al\mbox{.}(2024)]%
        {chan2024rq}
\bibfield{author}{\bibinfo{person}{Chi-Min Chan}, \bibinfo{person}{Chunpu Xu}, \bibinfo{person}{Ruibin Yuan}, \bibinfo{person}{Hongyin Luo}, \bibinfo{person}{Wei Xue}, \bibinfo{person}{Yike Guo}, {and} \bibinfo{person}{Jie Fu}.} \bibinfo{year}{2024}\natexlab{}.
\newblock \showarticletitle{Rq-rag: Learning to refine queries for retrieval augmented generation}.
\newblock \bibinfo{journal}{\emph{arXiv preprint arXiv:2404.00610}} (\bibinfo{year}{2024}).
\newblock


\bibitem[Chifu et~al\mbox{.}(2025)]%
        {chifu2025uncovering}
\bibfield{author}{\bibinfo{person}{Adrian-Gabriel Chifu}, \bibinfo{person}{S{\'e}bastien D{\'e}jean}, \bibinfo{person}{Moncef Garouani}, \bibinfo{person}{Josiane Mothe}, \bibinfo{person}{Di{\'e}go Ortiz}, {and} \bibinfo{person}{Md~Zia Ullah}.} \bibinfo{year}{2025}\natexlab{}.
\newblock \showarticletitle{Uncovering the Limitations of Query Performance Prediction: Failures, Insights, and Implications for Selective Query Processing}.
\newblock \bibinfo{journal}{\emph{ACM Transactions on Information Systems}} (\bibinfo{year}{2025}).
\newblock


\bibitem[Cronen-Townsend et~al\mbox{.}(2002)]%
        {cronen2002predicting}
\bibfield{author}{\bibinfo{person}{Steve Cronen-Townsend}, \bibinfo{person}{Yun Zhou}, {and} \bibinfo{person}{W~Bruce Croft}.} \bibinfo{year}{2002}\natexlab{}.
\newblock \showarticletitle{Predicting query performance}. In \bibinfo{booktitle}{\emph{Proceedings of the 25th annual international ACM SIGIR conference on Research and development in information retrieval}}. \bibinfo{pages}{299--306}.
\newblock


\bibitem[Cummins et~al\mbox{.}(2011)]%
        {cummins2011improved}
\bibfield{author}{\bibinfo{person}{Ronan Cummins}, \bibinfo{person}{Joemon Jose}, {and} \bibinfo{person}{Colm O'Riordan}.} \bibinfo{year}{2011}\natexlab{}.
\newblock \showarticletitle{Improved query performance prediction using standard deviation}. In \bibinfo{booktitle}{\emph{Proceedings of the 34th international ACM SIGIR conference on Research and development in Information Retrieval}}. \bibinfo{pages}{1089--1090}.
\newblock


\bibitem[Datta et~al\mbox{.}(2022)]%
        {datta2022pointwise}
\bibfield{author}{\bibinfo{person}{Suchana Datta}, \bibinfo{person}{Sean MacAvaney}, \bibinfo{person}{Debasis Ganguly}, {and} \bibinfo{person}{Derek Greene}.} \bibinfo{year}{2022}\natexlab{}.
\newblock \showarticletitle{A'Pointwise-Query, Listwise-Document'based Query Performance Prediction Approach}. In \bibinfo{booktitle}{\emph{Proceedings of the 45th International ACM SIGIR Conference on Research and Development in Information Retrieval}}. \bibinfo{pages}{2148--2153}.
\newblock


\bibitem[Dhole and Agichtein(2024)]%
        {dhole2024genqrensemble}
\bibfield{author}{\bibinfo{person}{Kaustubh~D Dhole} {and} \bibinfo{person}{Eugene Agichtein}.} \bibinfo{year}{2024}\natexlab{}.
\newblock \showarticletitle{Genqrensemble: Zero-shot llm ensemble prompting for generative query reformulation}. In \bibinfo{booktitle}{\emph{European Conference on Information Retrieval}}. Springer, \bibinfo{pages}{326--335}.
\newblock


\bibitem[Ebrahimi et~al\mbox{.}(2024)]%
        {ebrahimi2024estimating}
\bibfield{author}{\bibinfo{person}{Sajad Ebrahimi}, \bibinfo{person}{Maryam Khodabakhsh}, \bibinfo{person}{Negar Arabzadeh}, {and} \bibinfo{person}{Ebrahim Bagheri}.} \bibinfo{year}{2024}\natexlab{}.
\newblock \showarticletitle{Estimating query performance through rich contextualized query representations}. In \bibinfo{booktitle}{\emph{European Conference on Information Retrieval}}. Springer, \bibinfo{pages}{49--58}.
\newblock


\bibitem[Faggioli et~al\mbox{.}(2023)]%
        {faggioli2023geometric}
\bibfield{author}{\bibinfo{person}{Guglielmo Faggioli}, \bibinfo{person}{Nicola Ferro}, \bibinfo{person}{Cristina~Ioana Muntean}, \bibinfo{person}{Raffaele Perego}, {and} \bibinfo{person}{Nicola Tonellotto}.} \bibinfo{year}{2023}\natexlab{}.
\newblock \showarticletitle{A geometric framework for query performance prediction in conversational search}. In \bibinfo{booktitle}{\emph{Proceedings of the 46th International ACM SIGIR Conference on Research and Development in Information Retrieval}}. \bibinfo{pages}{1355--1365}.
\newblock


\bibitem[Faggioli et~al\mbox{.}(2022)]%
        {faggioli2022smare}
\bibfield{author}{\bibinfo{person}{Guglielmo Faggioli}, \bibinfo{person}{Oleg Zendel}, \bibinfo{person}{J~Shane Culpepper}, \bibinfo{person}{Nicola Ferro}, {and} \bibinfo{person}{Falk Scholer}.} \bibinfo{year}{2022}\natexlab{}.
\newblock \showarticletitle{sMARE: a new paradigm to evaluate and understand query performance prediction methods}.
\newblock \bibinfo{journal}{\emph{Information Retrieval Journal}} \bibinfo{volume}{25}, \bibinfo{number}{2} (\bibinfo{year}{2022}), \bibinfo{pages}{94--122}.
\newblock


\bibitem[Furnas et~al\mbox{.}(1987)]%
        {furnas1987vocabulary}
\bibfield{author}{\bibinfo{person}{George~W. Furnas}, \bibinfo{person}{Thomas~K. Landauer}, \bibinfo{person}{Louis~M. Gomez}, {and} \bibinfo{person}{Susan~T. Dumais}.} \bibinfo{year}{1987}\natexlab{}.
\newblock \showarticletitle{The vocabulary problem in human-system communication}.
\newblock \bibinfo{journal}{\emph{Commun. ACM}} \bibinfo{volume}{30}, \bibinfo{number}{11} (\bibinfo{year}{1987}), \bibinfo{pages}{964--971}.
\newblock


\bibitem[Ganguly et~al\mbox{.}(2022)]%
        {ganguly2022analysis}
\bibfield{author}{\bibinfo{person}{Debasis Ganguly}, \bibinfo{person}{Suchana Datta}, \bibinfo{person}{Mandar Mitra}, {and} \bibinfo{person}{Derek Greene}.} \bibinfo{year}{2022}\natexlab{}.
\newblock \showarticletitle{An analysis of variations in the effectiveness of query performance prediction}. In \bibinfo{booktitle}{\emph{European Conference on Information Retrieval}}. Springer, \bibinfo{pages}{215--229}.
\newblock


\bibitem[Gao et~al\mbox{.}(2023)]%
        {gao2023retrieval}
\bibfield{author}{\bibinfo{person}{Yunfan Gao}, \bibinfo{person}{Yun Xiong}, \bibinfo{person}{Xinyu Gao}, \bibinfo{person}{Kangxiang Jia}, \bibinfo{person}{Jinliu Pan}, \bibinfo{person}{Yuxi Bi}, \bibinfo{person}{Yixin Dai}, \bibinfo{person}{Jiawei Sun}, \bibinfo{person}{Haofen Wang}, {and} \bibinfo{person}{Haofen Wang}.} \bibinfo{year}{2023}\natexlab{}.
\newblock \showarticletitle{Retrieval-augmented generation for large language models: A survey}.
\newblock  \bibinfo{volume}{2}, \bibinfo{number}{1} (\bibinfo{year}{2023}).
\newblock


\bibitem[Hashemi et~al\mbox{.}(2019)]%
        {hashemi2019performance}
\bibfield{author}{\bibinfo{person}{Helia Hashemi}, \bibinfo{person}{Hamed Zamani}, {and} \bibinfo{person}{W~Bruce Croft}.} \bibinfo{year}{2019}\natexlab{}.
\newblock \showarticletitle{Performance Prediction for Non-Factoid Question Answering}. In \bibinfo{booktitle}{\emph{Proceedings of the 2019 ACM SIGIR International Conference on Theory of Information Retrieval}}. \bibinfo{pages}{55--58}.
\newblock


\bibitem[Hauff(2010)]%
        {hauff2010predicting}
\bibfield{author}{\bibinfo{person}{Claudia Hauff}.} \bibinfo{year}{2010}\natexlab{}.
\newblock \showarticletitle{Predicting the effectiveness of queries and retrieval systems}. In \bibinfo{booktitle}{\emph{SIGIR Forum}}, Vol.~\bibinfo{volume}{44}. \bibinfo{pages}{88}.
\newblock


\bibitem[Hauff et~al\mbox{.}(2009)]%
        {hauff2009combination}
\bibfield{author}{\bibinfo{person}{Claudia Hauff}, \bibinfo{person}{Leif Azzopardi}, {and} \bibinfo{person}{Djoerd Hiemstra}.} \bibinfo{year}{2009}\natexlab{}.
\newblock \showarticletitle{The combination and evaluation of query performance prediction methods}. In \bibinfo{booktitle}{\emph{European conference on information retrieval}}. Springer, \bibinfo{pages}{301--312}.
\newblock


\bibitem[Hauff et~al\mbox{.}(2010)]%
        {hauff2010query}
\bibfield{author}{\bibinfo{person}{Claudia Hauff}, \bibinfo{person}{Leif Azzopardi}, \bibinfo{person}{Djoerd Hiemstra}, {and} \bibinfo{person}{Franciska de Jong}.} \bibinfo{year}{2010}\natexlab{}.
\newblock \showarticletitle{Query performance prediction: Evaluation contrasted with effectiveness}. In \bibinfo{booktitle}{\emph{European Conference on Information Retrieval}}. Springer, \bibinfo{pages}{204--216}.
\newblock


\bibitem[He and Ounis(2004)]%
        {he2004inferring}
\bibfield{author}{\bibinfo{person}{Ben He} {and} \bibinfo{person}{Iadh Ounis}.} \bibinfo{year}{2004}\natexlab{}.
\newblock \showarticletitle{Inferring Query Performance Using Pre-retrieval Predictors.}. In \bibinfo{booktitle}{\emph{String Processing and Information Retrieval, 11th International Conference, {SPIRE} 2004, Padova, Italy, October 5-8, 2004, Proceedings}}. \bibinfo{pages}{43--54}.
\newblock
\href{https://doi.org/10.1007/978-3-540-30213-1\_5}{doi:\nolinkurl{10.1007/978-3-540-30213-1\_5}}


\bibitem[Hosseini et~al\mbox{.}(2024)]%
        {hosseini2024enhanced}
\bibfield{author}{\bibinfo{person}{Seyed~Mohammad Hosseini}, \bibinfo{person}{Negar Arabzadeh}, \bibinfo{person}{Morteza Zihayat}, {and} \bibinfo{person}{Ebrahim Bagheri}.} \bibinfo{year}{2024}\natexlab{}.
\newblock \showarticletitle{Enhanced retrieval effectiveness through selective query generation}. In \bibinfo{booktitle}{\emph{Proceedings of the 33rd ACM International Conference on Information and Knowledge Management}}. \bibinfo{pages}{3792--3796}.
\newblock


\bibitem[Jagerman et~al\mbox{.}(2023)]%
        {jagerman2023query}
\bibfield{author}{\bibinfo{person}{Rolf Jagerman}, \bibinfo{person}{Honglei Zhuang}, \bibinfo{person}{Zhen Qin}, \bibinfo{person}{Xuanhui Wang}, {and} \bibinfo{person}{Michael Bendersky}.} \bibinfo{year}{2023}\natexlab{}.
\newblock \showarticletitle{Query expansion by prompting large language models}.
\newblock \bibinfo{journal}{\emph{arXiv preprint arXiv:2305.03653}} (\bibinfo{year}{2023}).
\newblock


\bibitem[Kanoulas et~al\mbox{.}(2018)]%
        {kanoulas2018clef}
\bibfield{author}{\bibinfo{person}{Evangelos Kanoulas}, \bibinfo{person}{Dan Li}, \bibinfo{person}{Leif Azzopardi}, {and} \bibinfo{person}{Rene Spijker}.} \bibinfo{year}{2018}\natexlab{}.
\newblock \showarticletitle{CLEF 2018 technologically assisted reviews in empirical medicine overview}. In \bibinfo{booktitle}{\emph{CEUR workshop proceedings}}, Vol.~\bibinfo{volume}{2125}. CEUR-WS.
\newblock


\bibitem[Khodabakhsh et~al\mbox{.}(2024)]%
        {khodabakhsh2024bertpe}
\bibfield{author}{\bibinfo{person}{Maryam Khodabakhsh}, \bibinfo{person}{Fattane Zarrinkalam}, {and} \bibinfo{person}{Negar Arabzadeh}.} \bibinfo{year}{2024}\natexlab{}.
\newblock \showarticletitle{BertPE: a BERT-based pre-retrieval estimator for query performance prediction}. In \bibinfo{booktitle}{\emph{European Conference on Information Retrieval}}. Springer, \bibinfo{pages}{354--363}.
\newblock


\bibitem[Killingback and Zamani(2025)]%
        {killingback2025benchmarking}
\bibfield{author}{\bibinfo{person}{Julian Killingback} {and} \bibinfo{person}{Hamed Zamani}.} \bibinfo{year}{2025}\natexlab{}.
\newblock \showarticletitle{Benchmarking Information Retrieval Models on Complex Retrieval Tasks}.
\newblock \bibinfo{journal}{\emph{arXiv preprint arXiv:2509.07253}} (\bibinfo{year}{2025}).
\newblock


\bibitem[Krishna et~al\mbox{.}(2025)]%
        {krishna2025fact}
\bibfield{author}{\bibinfo{person}{Satyapriya Krishna}, \bibinfo{person}{Kalpesh Krishna}, \bibinfo{person}{Anhad Mohananey}, \bibinfo{person}{Steven Schwarcz}, \bibinfo{person}{Adam Stambler}, \bibinfo{person}{Shyam Upadhyay}, {and} \bibinfo{person}{Manaal Faruqui}.} \bibinfo{year}{2025}\natexlab{}.
\newblock \showarticletitle{Fact, fetch, and reason: A unified evaluation of retrieval-augmented generation}. In \bibinfo{booktitle}{\emph{Proceedings of the 2025 Conference of the Nations of the Americas Chapter of the Association for Computational Linguistics: Human Language Technologies (Volume 1: Long Papers)}}. \bibinfo{pages}{4745--4759}.
\newblock


\bibitem[Kwok(1996)]%
        {kwok1996new}
\bibfield{author}{\bibinfo{person}{Kuilam~L Kwok}.} \bibinfo{year}{1996}\natexlab{}.
\newblock \showarticletitle{A new method of weighting query terms for ad-hoc retrieval}. In \bibinfo{booktitle}{\emph{Proceedings of the 19th annual international ACM SIGIR conference on Research and development in information retrieval}}. \bibinfo{pages}{187--195}.
\newblock


\bibitem[Lewis et~al\mbox{.}(2020)]%
        {lewis2020retrieval}
\bibfield{author}{\bibinfo{person}{Patrick Lewis}, \bibinfo{person}{Ethan Perez}, \bibinfo{person}{Aleksandra Piktus}, \bibinfo{person}{Fabio Petroni}, \bibinfo{person}{Vladimir Karpukhin}, \bibinfo{person}{Naman Goyal}, \bibinfo{person}{Heinrich K{\"u}ttler}, \bibinfo{person}{Mike Lewis}, \bibinfo{person}{Wen-tau Yih}, \bibinfo{person}{Tim Rockt{\"a}schel}, {et~al\mbox{.}}} \bibinfo{year}{2020}\natexlab{}.
\newblock \showarticletitle{Retrieval-augmented generation for knowledge-intensive nlp tasks}.
\newblock \bibinfo{journal}{\emph{Advances in neural information processing systems}}  \bibinfo{volume}{33} (\bibinfo{year}{2020}), \bibinfo{pages}{9459--9474}.
\newblock


\bibitem[Li et~al\mbox{.}(2023)]%
        {li2023pseudo}
\bibfield{author}{\bibinfo{person}{Hang Li}, \bibinfo{person}{Ahmed Mourad}, \bibinfo{person}{Shengyao Zhuang}, \bibinfo{person}{Bevan Koopman}, {and} \bibinfo{person}{Guido Zuccon}.} \bibinfo{year}{2023}\natexlab{}.
\newblock \showarticletitle{Pseudo relevance feedback with deep language models and dense retrievers: Successes and pitfalls}.
\newblock \bibinfo{journal}{\emph{ACM Transactions on Information Systems}} \bibinfo{volume}{41}, \bibinfo{number}{3} (\bibinfo{year}{2023}).
\newblock


\bibitem[Ma et~al\mbox{.}(2023)]%
        {ma2023query}
\bibfield{author}{\bibinfo{person}{Xinbei Ma}, \bibinfo{person}{Yeyun Gong}, \bibinfo{person}{Pengcheng He}, \bibinfo{person}{Hai Zhao}, {and} \bibinfo{person}{Nan Duan}.} \bibinfo{year}{2023}\natexlab{}.
\newblock \showarticletitle{Query rewriting in retrieval-augmented large language models}. In \bibinfo{booktitle}{\emph{Proceedings of the 2023 Conference on Empirical Methods in Natural Language Processing}}. \bibinfo{pages}{5303--5315}.
\newblock


\bibitem[Meng et~al\mbox{.}(2023)]%
        {meng2023query}
\bibfield{author}{\bibinfo{person}{Chuan Meng}, \bibinfo{person}{Negar Arabzadeh}, \bibinfo{person}{Mohammad Aliannejadi}, {and} \bibinfo{person}{Maarten de Rijke}.} \bibinfo{year}{2023}\natexlab{}.
\newblock \showarticletitle{Query Performance Prediction: From Ad-hoc to Conversational Search}.
\newblock \bibinfo{journal}{\emph{arXiv preprint arXiv:2305.10923}} (\bibinfo{year}{2023}).
\newblock


\bibitem[Meng et~al\mbox{.}(2025)]%
        {meng2025query}
\bibfield{author}{\bibinfo{person}{Chuan Meng}, \bibinfo{person}{Negar Arabzadeh}, \bibinfo{person}{Arian Askari}, \bibinfo{person}{Mohammad Aliannejadi}, {and} \bibinfo{person}{Maarten~de Rijke}.} \bibinfo{year}{2025}\natexlab{}.
\newblock \showarticletitle{Query performance prediction using relevance judgments generated by large language models}.
\newblock \bibinfo{journal}{\emph{ACM Transactions on Information Systems}} \bibinfo{volume}{43}, \bibinfo{number}{4} (\bibinfo{year}{2025}), \bibinfo{pages}{1--35}.
\newblock


\bibitem[Miutra and Craswell(2018)]%
        {miutra2018introduction}
\bibfield{author}{\bibinfo{person}{Bhaskar Miutra} {and} \bibinfo{person}{Nick Craswell}.} \bibinfo{year}{2018}\natexlab{}.
\newblock \showarticletitle{An introduction to neural information retrieval}.
\newblock \bibinfo{journal}{\emph{Foundations and Trends{\^W} in Accounting}} \bibinfo{volume}{13}, \bibinfo{number}{1} (\bibinfo{year}{2018}), \bibinfo{pages}{1--126}.
\newblock


\bibitem[P{\'e}rez-Iglesias and Araujo(2010)]%
        {perez2010standard}
\bibfield{author}{\bibinfo{person}{Joaqu{\'\i}n P{\'e}rez-Iglesias} {and} \bibinfo{person}{Lourdes Araujo}.} \bibinfo{year}{2010}\natexlab{}.
\newblock \showarticletitle{Standard deviation as a query hardness estimator}. In \bibinfo{booktitle}{\emph{International Symposium on String Processing and Information Retrieval}}. Springer, \bibinfo{pages}{207--212}.
\newblock


\bibitem[Petroni et~al\mbox{.}(2024)]%
        {petroni2024ir}
\bibfield{author}{\bibinfo{person}{Fabio Petroni}, \bibinfo{person}{Federico Siciliano}, \bibinfo{person}{Fabrizio Silvestri}, {and} \bibinfo{person}{Giovanni Trappolini}.} \bibinfo{year}{2024}\natexlab{}.
\newblock \showarticletitle{IR-RAG@ SIGIR24: Information retrieval's role in RAG systems}. In \bibinfo{booktitle}{\emph{Proceedings of the 47th International ACM SIGIR Conference on Research and Development in Information Retrieval}}. \bibinfo{pages}{3036--3039}.
\newblock


\bibitem[Ponte and Croft(2017)]%
        {ponte2017language}
\bibfield{author}{\bibinfo{person}{Jay~M Ponte} {and} \bibinfo{person}{W~Bruce Croft}.} \bibinfo{year}{2017}\natexlab{}.
\newblock \showarticletitle{A language modeling approach to information retrieval}. In \bibinfo{booktitle}{\emph{ACM SIGIR Forum}}, Vol.~\bibinfo{volume}{51}. ACM New York, NY, USA, \bibinfo{pages}{202--208}.
\newblock


\bibitem[Pradeep et~al\mbox{.}(2025)]%
        {pradeep2025ragnarok}
\bibfield{author}{\bibinfo{person}{Ronak Pradeep}, \bibinfo{person}{Nandan Thakur}, \bibinfo{person}{Sahel Sharifymoghaddam}, \bibinfo{person}{Eric Zhang}, \bibinfo{person}{Ryan Nguyen}, \bibinfo{person}{Daniel Campos}, \bibinfo{person}{Nick Craswell}, {and} \bibinfo{person}{Jimmy Lin}.} \bibinfo{year}{2025}\natexlab{}.
\newblock \showarticletitle{Ragnar{\"o}k: A reusable RAG framework and baselines for TREC 2024 retrieval-augmented generation track}. In \bibinfo{booktitle}{\emph{European Conference on Information Retrieval}}. Springer.
\newblock


\bibitem[Pradeep et~al\mbox{.}(2024)]%
        {pradeep2024initial}
\bibfield{author}{\bibinfo{person}{Ronak Pradeep}, \bibinfo{person}{Nandan Thakur}, \bibinfo{person}{Shivani Upadhyay}, \bibinfo{person}{Daniel Campos}, \bibinfo{person}{Nick Craswell}, {and} \bibinfo{person}{Jimmy Lin}.} \bibinfo{year}{2024}\natexlab{}.
\newblock \showarticletitle{Initial nugget evaluation results for the trec 2024 rag track with the autonuggetizer framework}.
\newblock \bibinfo{journal}{\emph{arXiv preprint arXiv:2411.09607}} (\bibinfo{year}{2024}).
\newblock


\bibitem[Ray et~al\mbox{.}(2025)]%
        {ray2025metis}
\bibfield{author}{\bibinfo{person}{Siddhant Ray}, \bibinfo{person}{Rui Pan}, \bibinfo{person}{Zhuohan Gu}, \bibinfo{person}{Kuntai Du}, \bibinfo{person}{Shaoting Feng}, \bibinfo{person}{Ganesh Ananthanarayanan}, \bibinfo{person}{Ravi Netravali}, {and} \bibinfo{person}{Junchen Jiang}.} \bibinfo{year}{2025}\natexlab{}.
\newblock \showarticletitle{Metis: fast quality-aware rag systems with configuration adaptation}. In \bibinfo{booktitle}{\emph{Proceedings of the ACM SIGOPS 31st symposium on operating systems principles}}. \bibinfo{pages}{606--622}.
\newblock


\bibitem[Razavi et~al\mbox{.}(2025)]%
        {razavi2025benchmarking}
\bibfield{author}{\bibinfo{person}{Amirhossein Razavi}, \bibinfo{person}{Mina Soltangheis}, \bibinfo{person}{Negar Arabzadeh}, \bibinfo{person}{Sara Salamat}, \bibinfo{person}{Morteza Zihayat}, {and} \bibinfo{person}{Ebrahim Bagheri}.} \bibinfo{year}{2025}\natexlab{}.
\newblock \showarticletitle{Benchmarking prompt sensitivity in large language models}. In \bibinfo{booktitle}{\emph{European Conference on Information Retrieval}}. Springer, \bibinfo{pages}{303--313}.
\newblock


\bibitem[Roitman et~al\mbox{.}(2019)]%
        {roitman2019study}
\bibfield{author}{\bibinfo{person}{Haggai Roitman}, \bibinfo{person}{Shai Erera}, {and} \bibinfo{person}{Guy Feigenblat}.} \bibinfo{year}{2019}\natexlab{}.
\newblock \showarticletitle{A study of query performance prediction for answer quality determination}. In \bibinfo{booktitle}{\emph{Proceedings of the 2019 ACM SIGIR International Conference on Theory of Information Retrieval}}. \bibinfo{pages}{43--46}.
\newblock


\bibitem[Roitman et~al\mbox{.}(2017)]%
        {roitman2017robust}
\bibfield{author}{\bibinfo{person}{Haggai Roitman}, \bibinfo{person}{Shai Erera}, {and} \bibinfo{person}{Bar Weiner}.} \bibinfo{year}{2017}\natexlab{}.
\newblock \showarticletitle{Robust standard deviation estimation for query performance prediction}. In \bibinfo{booktitle}{\emph{Proceedings of the acm sigir international conference on theory of information retrieval}}. \bibinfo{pages}{245--248}.
\newblock


\bibitem[Salamat et~al\mbox{.}(2023)]%
        {salamat2023neural}
\bibfield{author}{\bibinfo{person}{Sara Salamat}, \bibinfo{person}{Negar Arabzadeh}, \bibinfo{person}{Shirin Seyedsalehi}, \bibinfo{person}{Amin Bigdeli}, \bibinfo{person}{Morteza Zihayat}, {and} \bibinfo{person}{Ebrahim Bagheri}.} \bibinfo{year}{2023}\natexlab{}.
\newblock \showarticletitle{Neural Disentanglement of Query Difficulty and Semantics}. In \bibinfo{booktitle}{\emph{CIKM}}. \bibinfo{pages}{4264--4268}.
\newblock


\bibitem[Salamat et~al\mbox{.}(2025)]%
        {salamat2025contrastive}
\bibfield{author}{\bibinfo{person}{Sara Salamat}, \bibinfo{person}{Negar Arabzadeh}, \bibinfo{person}{Shirin Seyedsalehi}, \bibinfo{person}{Amin Bigdeli}, \bibinfo{person}{Morteza Zihayat}, {and} \bibinfo{person}{Ebrahim Bagheri}.} \bibinfo{year}{2025}\natexlab{}.
\newblock \showarticletitle{A contrastive neural disentanglement approach for query performance prediction}.
\newblock \bibinfo{journal}{\emph{Machine Learning}} \bibinfo{volume}{114}, \bibinfo{number}{4} (\bibinfo{year}{2025}), \bibinfo{pages}{109}.
\newblock


\bibitem[Salemi and Zamani(2024)]%
        {salemi2024evaluating}
\bibfield{author}{\bibinfo{person}{Alireza Salemi} {and} \bibinfo{person}{Hamed Zamani}.} \bibinfo{year}{2024}\natexlab{}.
\newblock \showarticletitle{Evaluating retrieval quality in retrieval-augmented generation}. In \bibinfo{booktitle}{\emph{Proceedings of the 47th International ACM SIGIR Conference on Research and Development in Information Retrieval}}. \bibinfo{pages}{2395--2400}.
\newblock


\bibitem[Saleminezhad et~al\mbox{.}(2024)]%
        {saleminezhad2024context}
\bibfield{author}{\bibinfo{person}{Abbas Saleminezhad}, \bibinfo{person}{Negar Arabzadeh}, \bibinfo{person}{Soosan Beheshti}, {and} \bibinfo{person}{Ebrahim Bagheri}.} \bibinfo{year}{2024}\natexlab{}.
\newblock \showarticletitle{Context-aware query term difficulty estimation for performance prediction}. In \bibinfo{booktitle}{\emph{European Conference on Information Retrieval}}. Springer, \bibinfo{pages}{30--39}.
\newblock


\bibitem[Saleminezhad et~al\mbox{.}(2026a)]%
        {saleminezhad2026learning}
\bibfield{author}{\bibinfo{person}{Abbas Saleminezhad}, \bibinfo{person}{Negar Arabzadeh}, \bibinfo{person}{Soosan Beheshti}, {and} \bibinfo{person}{Ebrahim Bagheri}.} \bibinfo{year}{2026}\natexlab{a}.
\newblock \showarticletitle{Learning Context-aware Term Importance for Query Performance Prediction}.
\newblock \bibinfo{journal}{\emph{ACM Transactions on Intelligent Systems and Technology}} \bibinfo{volume}{17}, \bibinfo{number}{2} (\bibinfo{year}{2026}), \bibinfo{pages}{1--30}.
\newblock


\bibitem[Saleminezhad et~al\mbox{.}(2026b)]%
        {saleminezhad2026structure}
\bibfield{author}{\bibinfo{person}{Abbas Saleminezhad}, \bibinfo{person}{Negar Arabzadeh}, \bibinfo{person}{Seyed~Mohammad Hosseini}, \bibinfo{person}{Soosan Beheshti}, {and} \bibinfo{person}{Ebrahim Bagheri}.} \bibinfo{year}{2026}\natexlab{b}.
\newblock \showarticletitle{Structure-Aware Pre-retrieval Performance Prediction on Query Affinity Graphs}. In \bibinfo{booktitle}{\emph{European Conference on Information Retrieval}}. Springer, \bibinfo{pages}{547--556}.
\newblock


\bibitem[Santra et~al\mbox{.}(2026a)]%
        {santra2026beyond}
\bibfield{author}{\bibinfo{person}{Payel Santra}, \bibinfo{person}{Partha Basuchowdhuri}, {and} \bibinfo{person}{Debasis Ganguly}.} \bibinfo{year}{2026}\natexlab{a}.
\newblock \showarticletitle{Beyond Correlations: A Downstream Evaluation Framework for Query Performance Prediction}.
\newblock \bibinfo{journal}{\emph{arXiv preprint arXiv:2601.17339}} (\bibinfo{year}{2026}).
\newblock


\bibitem[Santra et~al\mbox{.}(2026b)]%
        {santra2026breakingflatgeneralisedquery}
\bibfield{author}{\bibinfo{person}{Payel Santra}, \bibinfo{person}{Partha Basuchowdhuri}, {and} \bibinfo{person}{Debasis Ganguly}.} \bibinfo{year}{2026}\natexlab{b}.
\newblock \bibinfo{title}{Breaking Flat: A Generalised Query Performance Prediction Evaluation Framework}.
\newblock
\showeprint[arxiv]{2601.17359}~[cs.IR]
\urldef\tempurl%
\url{https://arxiv.org/abs/2601.17359}
\showURL{%
\tempurl}


\bibitem[Scells et~al\mbox{.}(2018)]%
        {scells2018query}
\bibfield{author}{\bibinfo{person}{Harrisen Scells}, \bibinfo{person}{Leif Azzopardi}, \bibinfo{person}{Guido Zuccon}, {and} \bibinfo{person}{Bevan Koopman}.} \bibinfo{year}{2018}\natexlab{}.
\newblock \showarticletitle{Query variation performance prediction for systematic reviews}. In \bibinfo{booktitle}{\emph{The 41st International ACM SIGIR Conference on Research \& Development in Information Retrieval}}. \bibinfo{pages}{1089--1092}.
\newblock


\bibitem[Seo et~al\mbox{.}(2025)]%
        {seo2025newqueryexpansionapproach}
\bibfield{author}{\bibinfo{person}{Wonduk Seo}, \bibinfo{person}{Hyunjin An}, {and} \bibinfo{person}{Seunghyun Lee}.} \bibinfo{year}{2025}\natexlab{}.
\newblock \bibinfo{title}{A New Query Expansion Approach via Agent-Mediated Dialogic Inquiry}.
\newblock
\showeprint[arxiv]{2502.08557}~[cs.IR]
\urldef\tempurl%
\url{https://arxiv.org/abs/2502.08557}
\showURL{%
\tempurl}


\bibitem[Shtok et~al\mbox{.}(2010)]%
        {shtok2010using}
\bibfield{author}{\bibinfo{person}{Anna Shtok}, \bibinfo{person}{Oren Kurland}, {and} \bibinfo{person}{David Carmel}.} \bibinfo{year}{2010}\natexlab{}.
\newblock \showarticletitle{Using statistical decision theory and relevance models for query-performance prediction}. In \bibinfo{booktitle}{\emph{Proceedings of the 33rd international ACM SIGIR conference on Research and development in information retrieval}}. \bibinfo{pages}{259--266}.
\newblock


\bibitem[Shtok et~al\mbox{.}(2012)]%
        {shtok2012predicting}
\bibfield{author}{\bibinfo{person}{Anna Shtok}, \bibinfo{person}{Oren Kurland}, \bibinfo{person}{David Carmel}, \bibinfo{person}{Fiana Raiber}, {and} \bibinfo{person}{Gad Markovits}.} \bibinfo{year}{2012}\natexlab{}.
\newblock \showarticletitle{Predicting query performance by query-drift estimation}.
\newblock \bibinfo{journal}{\emph{ACM Transactions on Information Systems (TOIS)}} \bibinfo{volume}{30}, \bibinfo{number}{2} (\bibinfo{year}{2012}), \bibinfo{pages}{1--35}.
\newblock


\bibitem[Singh et~al\mbox{.}(2025)]%
        {singh2025agentic}
\bibfield{author}{\bibinfo{person}{Aditi Singh}, \bibinfo{person}{Abul Ehtesham}, \bibinfo{person}{Saket Kumar}, {and} \bibinfo{person}{Tala~Talaei Khoei}.} \bibinfo{year}{2025}\natexlab{}.
\newblock \showarticletitle{Agentic retrieval-augmented generation: A survey on agentic rag}.
\newblock \bibinfo{journal}{\emph{arXiv preprint arXiv:2501.09136}} (\bibinfo{year}{2025}).
\newblock


\bibitem[Tao and Wu(2014)]%
        {tao2014query}
\bibfield{author}{\bibinfo{person}{Yongquan Tao} {and} \bibinfo{person}{Shengli Wu}.} \bibinfo{year}{2014}\natexlab{}.
\newblock \showarticletitle{Query performance prediction by considering score magnitude and variance together}. In \bibinfo{booktitle}{\emph{Proceedings of the 23rd ACM international conference on conference on information and knowledge management}}. \bibinfo{pages}{1891--1894}.
\newblock


\bibitem[Thakur et~al\mbox{.}(2025)]%
        {thakur2025assessing}
\bibfield{author}{\bibinfo{person}{Nandan Thakur}, \bibinfo{person}{Ronak Pradeep}, \bibinfo{person}{Shivani Upadhyay}, \bibinfo{person}{Daniel Campos}, \bibinfo{person}{Nick Craswell}, \bibinfo{person}{Ian Soboroff}, \bibinfo{person}{Hoa~Trang Dang}, {and} \bibinfo{person}{Jimmy Lin}.} \bibinfo{year}{2025}\natexlab{}.
\newblock \showarticletitle{Assessing Support for the TREC 2024 RAG Track: A Large-Scale Comparative Study of LLM and Human Evaluations}. In \bibinfo{booktitle}{\emph{Proceedings of the 48th International ACM SIGIR Conference on Research and Development in Information Retrieval}}. \bibinfo{pages}{2759--2763}.
\newblock


\bibitem[Tian et~al\mbox{.}(2025a)]%
        {tian2025right}
\bibfield{author}{\bibinfo{person}{Fangzheng Tian}, \bibinfo{person}{Jinyuan Fang}, \bibinfo{person}{Debasis Ganguly}, \bibinfo{person}{Zaiqiao Meng}, {and} \bibinfo{person}{Craig Macdonald}.} \bibinfo{year}{2025}\natexlab{a}.
\newblock \showarticletitle{Am I on the Right Track? What Can Predicted Query Performance Tell Us about the Search Behaviour of Agentic RAG}.
\newblock  (\bibinfo{year}{2025}).
\newblock


\bibitem[Tian et~al\mbox{.}(2025b)]%
        {tian2025relevance}
\bibfield{author}{\bibinfo{person}{Fangzheng Tian}, \bibinfo{person}{Debasis Ganguly}, {and} \bibinfo{person}{Craig Macdonald}.} \bibinfo{year}{2025}\natexlab{b}.
\newblock \showarticletitle{Is Relevance Propagated from Retriever to Generator in RAG?}. In \bibinfo{booktitle}{\emph{European Conference on Information Retrieval}}. Springer, \bibinfo{pages}{32--48}.
\newblock


\bibitem[Voorhees(1993)]%
        {voorhees1993using}
\bibfield{author}{\bibinfo{person}{Ellen~M Voorhees}.} \bibinfo{year}{1993}\natexlab{}.
\newblock \showarticletitle{Using WordNet to disambiguate word senses for text retrieval}. In \bibinfo{booktitle}{\emph{Proceedings of the 16th annual international ACM SIGIR conference on Research and development in information retrieval}}. \bibinfo{pages}{171--180}.
\newblock


\bibitem[Wang et~al\mbox{.}(2023b)]%
        {wang2023query2doc}
\bibfield{author}{\bibinfo{person}{Liang Wang}, \bibinfo{person}{Nan Yang}, {and} \bibinfo{person}{Furu Wei}.} \bibinfo{year}{2023}\natexlab{b}.
\newblock \showarticletitle{Query2doc: Query expansion with large language models}.
\newblock \bibinfo{journal}{\emph{arXiv preprint arXiv:2303.07678}} (\bibinfo{year}{2023}).
\newblock


\bibitem[Wang et~al\mbox{.}(2023a)]%
        {wang2023generative}
\bibfield{author}{\bibinfo{person}{Xiao Wang}, \bibinfo{person}{Sean MacAvaney}, \bibinfo{person}{Craig Macdonald}, {and} \bibinfo{person}{Iadh Ounis}.} \bibinfo{year}{2023}\natexlab{a}.
\newblock \showarticletitle{Generative query reformulation for effective adhoc search}.
\newblock \bibinfo{journal}{\emph{arXiv preprint arXiv:2308.00415}} (\bibinfo{year}{2023}).
\newblock


\bibitem[Wang et~al\mbox{.}(2022)]%
        {wang2022self}
\bibfield{author}{\bibinfo{person}{Xuezhi Wang}, \bibinfo{person}{Jason Wei}, \bibinfo{person}{Dale Schuurmans}, \bibinfo{person}{Quoc Le}, \bibinfo{person}{Ed Chi}, \bibinfo{person}{Sharan Narang}, \bibinfo{person}{Aakanksha Chowdhery}, {and} \bibinfo{person}{Denny Zhou}.} \bibinfo{year}{2022}\natexlab{}.
\newblock \showarticletitle{Self-consistency improves chain of thought reasoning in language models}.
\newblock \bibinfo{journal}{\emph{arXiv preprint arXiv:2203.11171}} (\bibinfo{year}{2022}).
\newblock


\bibitem[Ye et~al\mbox{.}(2023)]%
        {ye2023enhancing}
\bibfield{author}{\bibinfo{person}{Fanghua Ye}, \bibinfo{person}{Meng Fang}, \bibinfo{person}{Shenghui Li}, {and} \bibinfo{person}{Emine Yilmaz}.} \bibinfo{year}{2023}\natexlab{}.
\newblock \showarticletitle{Enhancing conversational search: Large language model-aided informative query rewriting}.
\newblock \bibinfo{journal}{\emph{arXiv preprint arXiv:2310.09716}} (\bibinfo{year}{2023}).
\newblock


\bibitem[Zendel et~al\mbox{.}(2019)]%
        {DBLP:conf/sigir/ZendelSRKC19}
\bibfield{author}{\bibinfo{person}{Oleg Zendel}, \bibinfo{person}{Anna Shtok}, \bibinfo{person}{Fiana Raiber}, \bibinfo{person}{Oren Kurland}, {and} \bibinfo{person}{J.~Shane Culpepper}.} \bibinfo{year}{2019}\natexlab{}.
\newblock \showarticletitle{Information Needs, Queries, and Query Performance Prediction}. In \bibinfo{booktitle}{\emph{Proceedings of the 42nd International {ACM} {SIGIR} Conference on Research and Development in Information Retrieval, {SIGIR} 2019, Paris, France, July 21-25, 2019.}} \bibinfo{pages}{395--404}.
\newblock
\href{https://doi.org/10.1145/3331184.3331253}{doi:\nolinkurl{10.1145/3331184.3331253}}


\bibitem[Zhang et~al\mbox{.}(2024)]%
        {zhang2024exploring}
\bibfield{author}{\bibinfo{person}{Le Zhang}, \bibinfo{person}{Yihong Wu}, \bibinfo{person}{Qian Yang}, {and} \bibinfo{person}{Jian-Yun Nie}.} \bibinfo{year}{2024}\natexlab{}.
\newblock \showarticletitle{Exploring the best practices of query expansion with large language models}.
\newblock \bibinfo{journal}{\emph{arXiv preprint arXiv:2401.06311}} (\bibinfo{year}{2024}).
\newblock


\bibitem[Zhao et~al\mbox{.}(2024)]%
        {zhao2024longrag}
\bibfield{author}{\bibinfo{person}{Qingfei Zhao}, \bibinfo{person}{Ruobing Wang}, \bibinfo{person}{Yukuo Cen}, \bibinfo{person}{Daren Zha}, \bibinfo{person}{Shicheng Tan}, \bibinfo{person}{Yuxiao Dong}, {and} \bibinfo{person}{Jie Tang}.} \bibinfo{year}{2024}\natexlab{}.
\newblock \showarticletitle{Longrag: A dual-perspective retrieval-augmented generation paradigm for long-context question answering}.
\newblock  (\bibinfo{year}{2024}).
\newblock


\bibitem[Zhao et~al\mbox{.}(2008)]%
        {zhao2008effective}
\bibfield{author}{\bibinfo{person}{Ying Zhao}, \bibinfo{person}{Falk Scholer}, {and} \bibinfo{person}{Yohannes Tsegay}.} \bibinfo{year}{2008}\natexlab{}.
\newblock \showarticletitle{Effective Pre-retrieval Query Performance Prediction Using Similarity and Variability Evidence}. In \bibinfo{booktitle}{\emph{Advances in Information Retrieval , 30th European Conference on {IR} Research, {ECIR} 2008, Glasgow, UK, March 30-April 3, 2008. Proceedings}}. \bibinfo{pages}{52--64}.
\newblock


\bibitem[Zhou and Croft(2007)]%
        {zhou2007query}
\bibfield{author}{\bibinfo{person}{Yun Zhou} {and} \bibinfo{person}{W~Bruce Croft}.} \bibinfo{year}{2007}\natexlab{}.
\newblock \showarticletitle{Query performance prediction in web search environments}. In \bibinfo{booktitle}{\emph{Proceedings of the 30th annual international ACM SIGIR conference on Research and development in information retrieval}}. \bibinfo{pages}{543--550}.
\newblock


\end{thebibliography}
